\documentclass[apsrev4-1,prl,twocolumn,times,amsmath,amssymb,superscriptaddress,floatfix,longbibliography]{revtex4-1}

\usepackage{times}
\usepackage{float}
\usepackage{amsmath}
\usepackage{amsthm}
\usepackage[colorlinks,bookmarks=false,citecolor=blue,linkcolor=red,urlcolor=blue]{hyperref}
\usepackage{amssymb}
\usepackage{amsbsy}
\usepackage{graphicx}
\usepackage{pstricks}
\usepackage{color}
\usepackage{cancel}
\usepackage{subfigure} 
\usepackage{enumerate}
\usepackage{mathrsfs} 
\usepackage{multirow} 
\usepackage{wasysym}
\usepackage{sidecap}

\begin{document}
\title{Quantum spin ice with frustrated transverse exchange :  
        from $\pi$--flux phase\\ to nematic quantum spin liquid}


\author{Owen Benton}
\affiliation{RIKEN Center for Emergent Matter Science (CEMS), Wako, Saitama, 351-0198,
Japan}

\author{L. D. C. Jaubert}
\affiliation{CNRS, Universit\'e de Bordeaux, LOMA, UMR 5798, 33400 Talence, France}

\author{Rajiv Singh}
\affiliation{Department of Physics, University of California, Davis, California 95616, USA}

\author{Jaan Oitmaa}
\affiliation{School of Physics, The University of New South Wales, Sydney 2052, Australia}

\author{Nic Shannon}
\affiliation{Okinawa Institute of Science and Technology Graduate University,
Onna-son, Okinawa 904-0495, Japan}


\begin{abstract}
Quantum spin ice materials, pyrochlore magnets with competing Ising 
and transverse exchange interactions, have been widely discussed 
as candidates for a quantum spin--liquid ground state. 
Here, motivated by quantum chemical calculations for Pr pyrochlores, 
we present the results of a study for frustrated transverse exchange. 
Using a combination of variational calculations, 
exact diagonalisation, numerical linked-cluster and series expansions, we find 
that the previously-studied $U(1)$ quantum spin liquid, 
in its $\pi$-flux phase, transforms into a nematic quantum spin liquid at 
a high--symmetry, $SU(2)$ point.    
\end{abstract}

\maketitle
%

Pyrochlore magnets have proved an exceptionally rich source 
of new phenomena \cite{gardner10,hallas-arXiv}, including the 
classical spin liquid ``spin ice'' \cite{bramwell01,castelnovo12}, celebrated 
for its magnetic monopole excitations \cite{castelnovo08}.
Pyrochlore materials also stand at the forefront of the search for 
quantum spin liquids (QSL), massively--entangled quantum 
phases of matter, which provide accessible examples
of exotic, topological \mbox{(quasi--)particles} 
previously studied in high--energy physics 
\cite{balents10, savary17-RPP80, zhou17, norman16}.
In particular, the quantum analogue of spin ice has been shown to 
support a three--dimensional QSL with fractional excitations, 
described by a $U(1)$ lattice gauge theory \cite{hermele04,banerjee08,shannon12,benton12,kato15,huang-arXiv}, 
and has been vigorously pursued in experiment 
\cite{ross11-PRX1,Kimura2013,petit16-PRB94,wen17,martin17,sibille-arXiv}.


Exciting as these developments 
are, the range of outcomes in experiment 
remains far broader than 
predicted by theory \cite{gardner10,hallas-arXiv}.
Encouragingly, studies of more general pyrochlore--lattice 
models, in their classical limit, reveal a variety of new ordered 
and spin--liquid phases, which may provide insight into 
experiments carried out at finite temperature 
\cite{Onoda2011a,benton16-NatCommun7,petit16-PRB94,yan17,taillefumier17}.   
However, to date, very little is known about the ground state of 
even the simplest model of a quantum spin ice for frustrated 
transverse exchange, where quantum Monte Carlo simulation 
fails \cite{lee12,chen17-PRB96A}.
And, since microscopic estimates for Pr--based pyrochlore  
magnets have predicted frustrated interactions \cite{Onoda2011a}, 
this is a question of both fundamental and experimental interest.

\begin{figure}
\centering
\includegraphics[width=0.475\columnwidth]{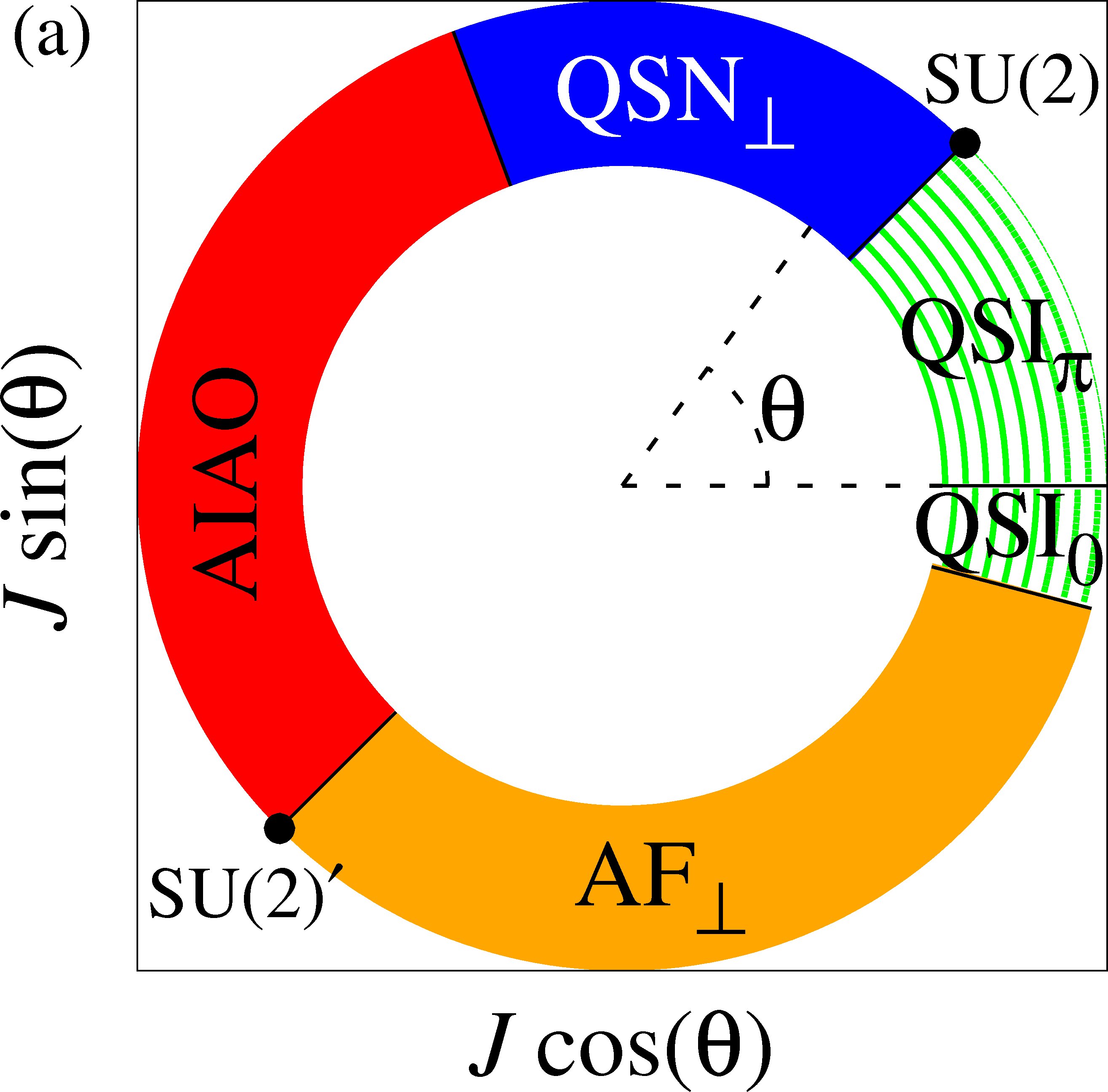}
\;
\includegraphics[width=0.46\columnwidth]{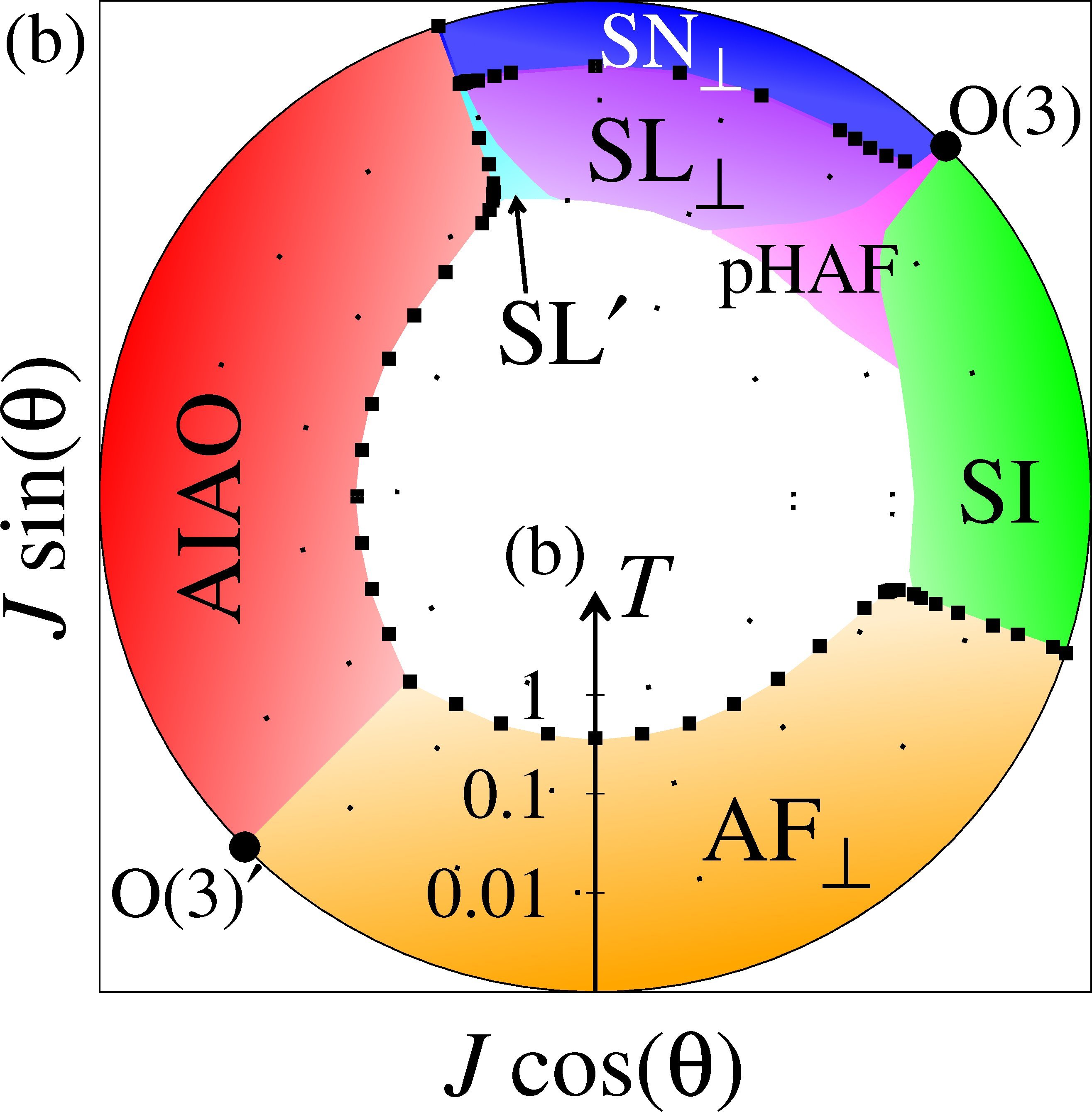}
\caption{
Phase diagram of the quantum spin ice model $\mathcal{H}_{\sf XXZ}$ 
[Eq.~(\ref{eq:Hxxz}), Eq.~(\ref{eq:J.theta})]:
(a)~Quantum phase diagram 
found in cluster--variational (cVAR) 
calculations for $T=0$.
Two quantum spin liquids (QSL) 
descended from spin ice, QSI$_0$ and QSI$_\pi$, 
compete with easy--plane antiferromagnetic order (AF$_\perp$); 
all--in all-out magnetic order (AIAO); 
and a nematic QSL (QSN$_\perp$).
QSI$_\pi$ and QSN$_\perp$ are 
connected by a point with $SU(2)$ symmetry.  
(b)~Phase diagram found in classical Monte Carlo 
simulations for $T>0$ [cf. Ref.~\onlinecite{taillefumier17}].
Three spin liquids; spin ice (SI); 
the easy--plane spin liquid (SL$_\perp$); 
and a pseudo--Heisenberg antiferromagnet (pHAF);
compete with a nematic spin liquid (SN$_\perp$); and 
AF$_\perp$ and AIAO order.  
An additional disordered regime
(SL$^\prime$) shares the correlations of SL$_{\perp}$ and AIAO. 
Details of simulation methods are given in the 
Supplemental Materials.
\label{fig:phase.diagram}
}
\end{figure}

In this Letter we address the fate of the QSL 
in a quantum spin ice with frustrated transverse exchange.
We find that the $U(1)$ QSL, in its $\pi$-flux phase \cite{lee12}, 
gives way to a new, nematic QSL, 
at a special $SU(2)$--symmetry point in parameter space.
Evidence in support of this claim is taken from exact diagonalisation (ED); 
a cluster--based mean--field theory (CMFT); 
cluster--based variational calculations (cVAR);
and an exact, variational argument at the $SU(2)$  point.
Further evidence for the growth of nematic correlations,  
and of an unusual scaling of heat capacity at high 
temperature, are presented through numerical linked--cluster 
expansion (NLCE) and high--temperature series expansion (HTE) 
calculations.
These results, summarised in Fig.~\ref{fig:phase.diagram}, 
provide a concrete example of a nematic quantum spin liquid  \cite{grover10}, 
in three dimensions, and confirm that even the simplest models of pyrochlore 
magnets can support a range of different QSL ground states.


The model we consider is the spin--$1/2$ XXZ Hamiltonian on 
the pyrochlore lattice 
\begin{eqnarray}
\mathcal{H}_{\sf XXZ} 
   = \sum_{\langle ij \rangle} 
   \left[
    J_{zz} {\sf S}^z_i {\sf S}^z_j
    - J_\pm \left(
    {\sf S}^+_i {\sf S}^-_j 
    + {\sf S}^+_j {\sf S}^-_i 
    \right) 
   \right]
\label{eq:Hxxz}
\end{eqnarray}
where spin coordinates are defined in a local coordinate
frame such that the $z$-axis of spin space is aligned with
a local $C_3$ axis \cite{ross11-PRX1,yan17}. 
Eq.~(\ref{eq:Hxxz}) can be derived from atomic models of pyrochlore 
oxides \cite{molavian07, Onoda2010, Onoda2011a} and, for
$J_{zz} \gg J_\pm > 0$, has been extensively studied as a minimal model 
of a quantum spin ice \cite{hermele04,banerjee08,shannon12,benton12,kato15,huang-arXiv,
lee12,chen17-PRB96A,savary12-PRL108,Hao2014,
mcclarty15,gingras14,chen16,shannon-book-chapter,savary17-PRL118,chen17-PRB96B}.
Since we are concerned with both positive and negative 
signs of interaction, it is convenient to write
\begin{eqnarray}
   J_{zz} = J \cos \theta  \quad ,  \quad
   J_\pm = - \frac{J}{2} \sin \theta  \; .
   \label{eq:J.theta}
\end{eqnarray}
At the special points $\theta = \pi/4$, and $\theta = -3\pi/4$, 
$\mathcal{H}_{\sf XXZ}$ is equivalent to a Heisenberg model, 
and has an $SU(2)$ symmetry.


For unfrustrated interactions, \mbox{$-\pi/2 < \theta < 0$}, 
$\mathcal{H}_{\sf XXZ}$~[Eq.~(\ref{eq:Hxxz})]
is accessible to quantum Monte Carlo (QMC) simulation.
In this case, for $\theta \lesssim 0$, 
the ground--state
is known to be a $U(1)$ QSL  (QSI$_0$), 
giving way to an easy--plane antiferromagnet (AF$_\perp$) 
for $\theta < -0.05$ 
\cite{banerjee08,shannon12,kato15}.
Perturbative arguments imply that the $U(1)$ QSL
should also survive for frustrated interactions, $\theta \gtrsim 0$ 
\cite{hermele04}.
In this case the $U(1)$ QSL enters a ``$\pi$--flux phase'' (QSI$_\pi$), 
in which its topological, spinon excitations have a modified 
dispersion, due to a fractionalisation of translational 
symmetry \cite{lee12,chen17-PRB96A}.  
Classical Monte Carlo simulations 
suggest that $\mathcal{H}_{\sf XXZ}$ remains in a spin 
liquid state for \mbox{$0 < \theta < 1.89$}, but that this spin liquid 
changes its character traversing the 
high--symmetry point $\theta = \pi/4$ \cite{taillefumier17} 
--- cf. Fig.~\ref{fig:phase.diagram}(b).
The fate of the quantum, $\pi$--flux ground state, however, 
remains unknown.


{\it CMFT}-- 
In order to shed light on this question, we first explore the ground 
state of $\mathcal{H}_{\sf XXZ}$~[Eq.~(\ref{eq:Hxxz})] within  
an approach based on cluster mean field theory (CMFT).
CMFT consists in breaking the lattice up into finite clusters
and treating the interactions within each cluster exactly, 
while those between clusters are treated at a mean--field 
level \cite{garcia00, garcia01, shannon02, yamamoto14, javanparast15}.
The geometry of the pyrochlore lattice permits degenerate CMFT solutions,
with translational symmetry restored, in contrast to some previous approaches 
(see e.g. \cite{yamamoto15-PRL114}), allowing us to treat spin--liquid states.


We start by dividing the pyrochlore lattice into two sublattices of tetrahedra, 
`A' and `B', and writing the wave function as a product over A--sublattice tetrahedra
\begin{eqnarray}
   | \psi^{\sf CMFT}  (\{ {\bf h} \}) \rangle = \Pi_{t \in A}  |\phi_t(\{{\bf h}\})\rangle  \; ,
\label{eq:psi.CMFT}
\end{eqnarray}
where $|\phi_t(\{{\bf h}\})\rangle$ is defined as the ground state of an 
auxiliary Hamiltonian on tetrahedron $t$
\begin{eqnarray}
   \mathcal{H}'(t) = \mathcal{H}_{\sf XXZ} (t) - \sum_{i \in t} {\bf h}_i \cdot {\sf S}_i \; .
\label{eq:aux_H}
\end{eqnarray}
Correlations within B--sublattice tetrahedra are treated at a mean--field level, 
through the self--consistently determined field $\{{\bf h}\}$
\begin{eqnarray}
   \mathcal{H}'(t) |\phi_t(\{{\bf h}\})\rangle = \epsilon_t  |\phi_t(\{{\bf h}\})\rangle \; ,
\label{eq:aux_H_eigval}
\end{eqnarray}
with the optimal values of  $\{ {\bf h} \}$ found variationally, 
by minimising 
\begin{eqnarray}
   E_{\sf CMFT} = \langle \psi^{\sf CMFT}  (\{ {\bf h} \})| 
   \mathcal{H}_{\sf XXZ} | \psi^{\sf CMFT}  (\{ {\bf h} \}) \rangle \; .
\label{eq:Evar_CMFT}
\end{eqnarray}
The corner--sharing geometry of the pyrochlore lattice permits 
solutions for a single tetrahedron to be connected in many 
different ways (cf. {\it ``lego--brick rules''} in \cite{yan17}).   
For this reason the solution for $\{ {\bf h} \}$, and the corresponding 
wave function $| \psi^{\sf CMFT}  (\{ {\bf h} \}) \rangle$, 
encompass disordered as well as ordered states.   


We find four kinds of optimal solutions for the fields ${\bf h}_i$,  
each corresponding to a different region of the phase diagram 
Fig.~\ref{fig:phase.diagram}(a).  
For 
$5\pi/4  < \theta < 1.927$,  
the optimal solution has 
${\bf h}_i=h \hat{{\bf z}}$
on all sites, and corresponds to all--in, all--out (AIAO) order.
Meanwhile, for $-\frac{3\pi}{4}<\theta\lesssim-0.256$ 
fields ${\bf h_i}$ are globally ordered in the local $xy$ plane, with (e.g.)
 ${\bf h}_i=h \hat{{\bf x}}$. 
 This is the easy--plane antiferromagnet, AF$_{\perp}$.


For $-0.256\lesssim\theta<\frac{\pi}{4}$ the optimal solutions are spin--ice--like.
The fields ${\bf h}_i$ have the form ${\bf h}_i=\sigma_i h \hat{{\bf z}}$ where $\sigma_i=\pm1$.
The minimum value of $E_{\sf CMFT}$ is attained by any configuration of $\sigma_i$
with two `+' signs and two `-' signs on every tetrahedron of the lattice.
It is known from perturbative arguments that quantum tunnelling 
between spin--ice configurations gives rise to two distinct $U(1)$ QSL, 
depending on the sign of $J_\pm$ \cite{lee12,chen17-PRB96A}.
These two phases, QSI$_0$ and QSI$_\pi$, cannot be distinguished within CMFT, 
but do appear as distinct phases in more sophsticated variational calculations, discussed below.


For $\frac{\pi}{4}\lesssim\theta<1.927$ the optimal solutions are similar to the
spin--ice case but now have the fields ${\bf h}_i$ lying in the $xy$ plane,
in a collinear fashion, e.g. ${\bf h}_i=\sigma_i h \hat{{\bf x}}$.
Once again $E_{\sf CMFT}$ is minimized by any configuration of $\sigma_i$
with two `+' signs and two `-' signs on every tetrahedron.
Since these $\sigma_i$ are disordered,  
the resulting state does not possess any conventional magnetic order.
None the less, the selection of a global axis in the $xy$ plane implies that  
it breaks the $U(1)$ spin--rotation symmetry of Eq.~(\ref{eq:Hxxz}).
And this is reflected in a finite value of the spin--nematic 
order parameter 
\begin{eqnarray}
\mathcal{Q}_\perp = \bigg\langle
\frac{1}{3N}
\sum_{\langle ij \rangle} 
\begin{pmatrix}
{\sf S}^x_i {\sf S}^x_j - {\sf S}^y_i {\sf S}^y_j \\
{\sf S}^x_i {\sf S}^y_j + {\sf S}^y_i {\sf S}^x_j 
\end{pmatrix} \bigg\rangle \; ,
\label{eq:Qperp}
\end{eqnarray}
defined on the bonds $\langle ij \rangle$ of the pyrochlore 
lattice \cite{taillefumier17}.


{\it cVAR--} 
The CMFT wave function, Eq.~(\ref{eq:psi.CMFT}), is entangled at the 
level of a single tetrahedron, and can describe disordered 
as well as ordered states.
But, it cannot capture the long--range 
entanglement of a QSL.
For this reason, distinguishing the quantum ground states 
of Eq.~(\ref{eq:Hxxz}) requires going
beyond mean--field theory.
To this end, we now introduce a cluster--variational
(cVAR) approach, based on a coherent superposition of the 
degenerate ground states found in CMFT.
We apply this method to the case where CMFT 
predicts spin--nematic order, finding that quantum fluctuations beyond CMFT 
lead to a $U(1)$ QSL, which retains spin--nematic order.
Further details of the cVAR approach, including its application to 
the two QSLs descended from spin ice, QSI$_0$ and QSI$_\pi$, 
can be found in the Supplementary Materials.


%
We take as a starting point the CMFT ansatz for a spin--nematic state 
with axis of collinearity ${\bf h} \parallel \hat{{\bf x}}$.
A superposition of such wave functions can be written as
\begin{eqnarray}
|\varphi\rangle= \sum_{\{ \sigma \}} a_{\{\sigma\}} | \psi^{\sf CMFT} (h \sigma_i \hat{\bf x}) \rangle \; ,
\label{eq:mf_superpose}
\end{eqnarray}
where the sum runs over all Ising configurations $\{ \sigma \}$ with two `+'
and two `-' on every tetrahedron.
The complex coefficients $a_{\{\sigma\}}$ are the variational 
parameters with which we can further optimize the energy
\begin{eqnarray}
E_{\sf cVAR} = \frac{\langle \varphi | \mathcal{H}_{\sf XXZ} | \varphi\rangle}{\langle \varphi | \varphi \rangle } \; .
\label{eq:newvar}
\end{eqnarray}


The wavefunctions $| \psi^{\sf CMFT} (h \sigma_i \hat{\bf x}) \rangle$ labelled by different 
Ising configurations  $\{ \sigma \}$ are not generally orthogonal. 
The overlap between different mean--field solutions can be parameterised by 
a dimensionless quantity $\mu(\theta)$, with $|\mu(\theta)|<1$
The overlap between two optimized CMFT wavefunctions is then $\sim \mu^{N_{\sf diff}}$ where 
$N_{\sf diff}$ is the number of `A' tetrahedra on which the 
arrangement of $\sigma_i$ differs between the
two.
Using this fact we can expand both numerator and denominator 
of Eq.~(\ref{eq:newvar}) in powers of $\mu$.
When $|\mu|\ll 1$, we may justify keeping only the leading term 
which reduces Eq.~(\ref{eq:newvar})
to 
\begin{eqnarray}
E_{\sf cVAR}
   \approx E_{\sf CMFT} 
 + \sum_{\{ \sigma\}, \{ \sigma' \}}M_{\{\sigma\} \{\sigma'\}} a^{\ast}_{\{\sigma'\}} 
a_{\{\sigma\}} \quad
\label{eq:newvar-expanded}
\end{eqnarray}
where the matrix element $M_{\{\sigma\} \{\sigma'\}} $ is a constant $ \propto \mu^2$
for two configurations connected by reversing the signs of $\sigma_i$ around a single 
hexagonal plaquette, and zero otherwise. 


It follows that minimizing the variational energy in Eq.~(\ref{eq:newvar-expanded})
is equivalent to finding the ground state of the ring--exchange 
Hamiltonian studied using QMC in \cite{shannon12}, where the outcome is a $U(1)$ QSL.
This implies that 
the optimal superposition of CMFT wave functions, $|\phi \rangle$ [Eq.~(\ref{eq:mf_superpose})], 
is also a $U(1)$ QSL.
Moreover, since each of these mean--field solutions has the same value 
of $\mathcal{Q}_\perp$ [Eq.~(\ref{eq:Qperp})], this $U(1)$ QSL
retains the spin--nematic order found in CMFT.
Following \cite{grover10}, we dub this phase 
a {\it ``nematic quantum spin liquid''}, and denote it QSN$_\perp$ in 
Fig.~\ref{fig:phase.diagram}(a).
Evaluating $\mu(\theta)$ numerically, we find \mbox{$|\mu(\theta)| < 0.5$} for all relevant 
parameters,  
with $|\mu(\theta)| \to 0.15$ approaching AIAO order.
This suggests that the perturbative expansion of Eq.~(\ref{eq:newvar}) is justified.


{\it Further support for nematic order}-- 
We now provide two further arguments, completely independent of the 
cVAR approach, which support the existence of spin--nematic order.


The first argument is based on approaching
the $SU(2)$  point $\theta=\frac{\pi}{4}$ from the small
$\theta$ side.
For small $\theta>0$ the ground state is
the $\pi$-flux $U(1)$ QSL \cite{hermele04, lee12, chen17-PRB96A}
(QSI$_\pi$ in Fig.~\ref{fig:phase.diagram}(a)).
Gauge Mean Field Theory predicts this state to be stable up to 
$\theta\approx1.12$, well beyond the $SU(2)$  point.
However, we show below that if 
QSI$_\pi$ is stable up to the $SU(2)$  point,
it must at that point become unstable to nematicity.


To see this, we observe that an appropriate trial wavefunction
for the spin--nematic phase can be generated by taking a
ground state wavefunction from within the 
QSI$_\pi$ phase and acting on it with global spin rotations:
\begin{eqnarray}
|\text{nem}(\psi)\rangle
=\mathcal{R}_z(\psi)\mathcal{R}_y\left(\frac{\pi}{2} \right)|\text{QSI}_{\pi}\rangle \; ,
\label{eq:nem_wavefunction}
\end{eqnarray}
where $\mathcal{R}_{\alpha}(\phi)$ denotes a global rotation by an angle
$\phi$, around the $\alpha$ axis of spin space.
The wavefunction $|\text{nem}(\psi)\rangle$ generically supports a finite
value of the nematic bond order parameter $\mathcal{Q}_{\perp}$
[Eq.~(\ref{eq:Qperp})], with all dipolar expectation values vanishing. 
The angle $\psi$ parameterises the direction of $\mathcal{Q}_{\perp}$ 
in the nematic state.


Eq.~(\ref{eq:nem_wavefunction}) links a wavefunction describing the spin nematic phase
with a wavefunction describing QSI$_\pi$, using global spin rotations.
These spin rotations become symmetries of the model at the $SU(2)$  point
$\theta=\frac{\pi}{4}$.
Thus, if QSI$_\pi$ is stable up to the $SU(2)$  point, the energy
gap between this state and the spin nematic must vanish there, indicating an incipient
instability to nematicity. 
It follows that the resulting spin--nematic state inherits both the gauge symmetry
and the fractionalised translational symmetry of QSI$_\pi$.


The argument above cannot, however, rule out the possibility that some other ground state may take over
from QSI$_\pi$ before $\theta=\frac{\pi}{4}$, and have yet lower energy. 
Such alternative competing ground states around $\theta=\frac{\pi}{4}$ could include various
dimer-ordered \cite{harris91, berg03, tsunetsugu01-JPSJ70, tsunetsugu01-PRB65, moessner06} 
and spin liquid \cite{canals98, canals00, kim08, burnell09} ground states suggested for the 
Heisenberg model previously.
It is useful therefore to have an alternative way to establish nematic
order.
This is provided by considering the excitations of the AIAO ordered phase
found for $J^{zz} < 0$ [cf. Fig.~\ref{fig:phase.diagram}(a)].


In the AIAO phase the ground state wavefunction is simply the polarized state with
maximum total $S^z$.
Since total $S^z$ is a conserved quantity, the excitations of the AIAO phase can be
labelled by the number of spin flips, $\delta S^z$, relative to the AIAO
ground state.


\begin{figure}
\centering
\includegraphics[width=\columnwidth]{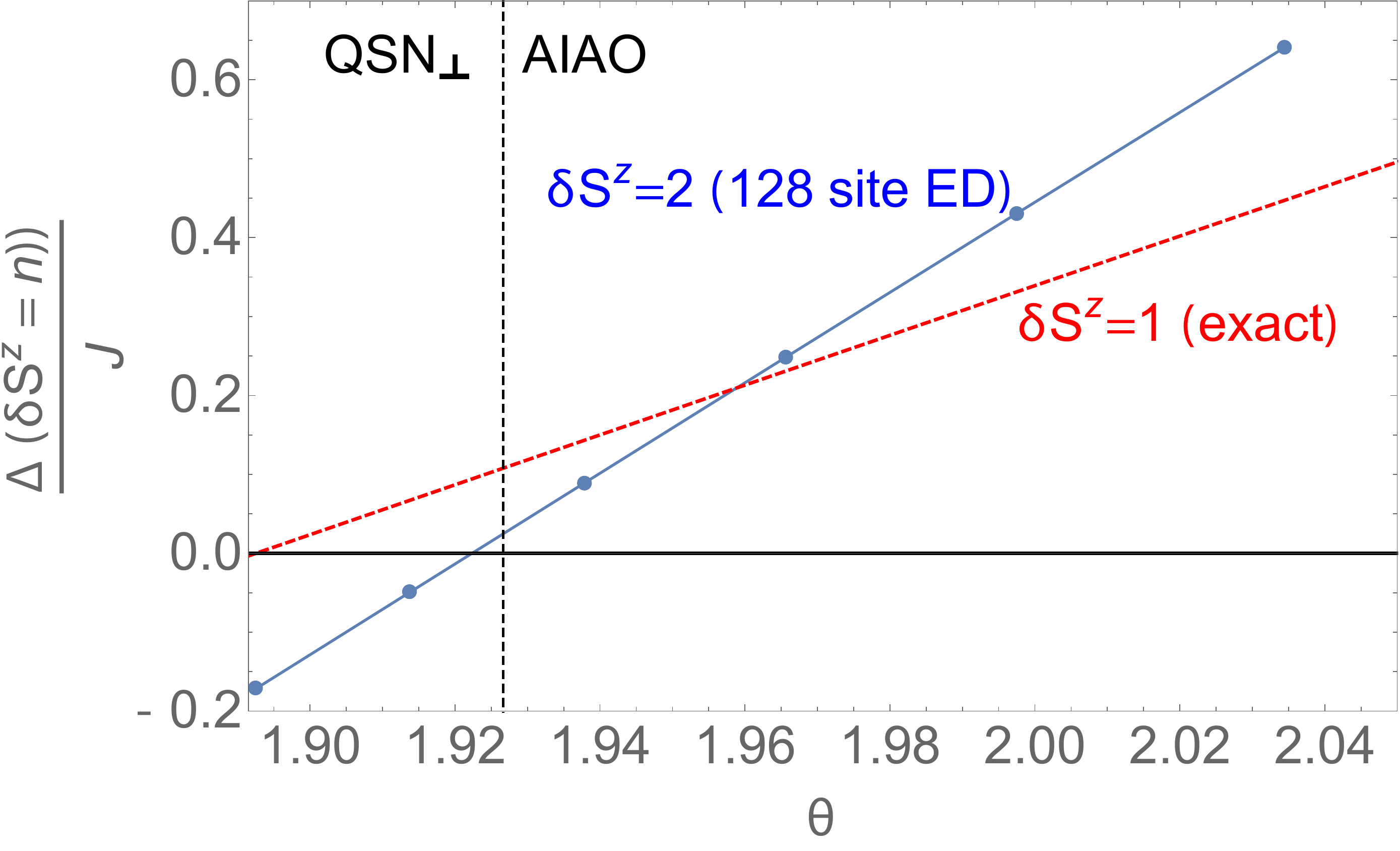}
\caption{Condensation of two--magnon bound states within the AIAO phase,
indicating the onset of spin--nematic order.
The gaps to the lowest 1--magnon and 2--magnon excitations of the 
AIAO phase, $\Delta(\delta S^z=1)$ and $\Delta(\delta S^z=2)$, are 
shown as a function of the Hamiltonian parameter $\theta$ [Eq.~(\ref{eq:J.theta})].  
For $\delta S^z=1$ the gap  has been calculated exactly, while for 
$\delta S^z=2$ it has been estimated numerically for a cluster of 128 spins.   
As $\theta$ decreases towards $\theta=1.92$
the two--magnon state comes below the one magnon state and then
crosses the AIAO state energy.
The condensation of two--magnon bound states is a clear indication
of incipient nematic order \cite{shannon06}.
This estimate of the phase boundary between QSN$_\perp$ and
AIAO, $\theta\approx1.924$, is very close to that found in cVAR  
$\theta\approx1.927$ [cf. Fig.~\ref{fig:phase.diagram}(a)], 
shown here as a vertical dashed line.
}
\label{fig:boundstate}
\end{figure}


Starting from the AIAO state, an instability to a conventional XY ordered
state would be indicated by the softening of a $\delta S^z=1$ excitation-
i.e. a magnon.
An instability to nematic order, by contrast, would be indicated by 
the softening of a $\delta S^z=2$ excitation: a two-magnon
bound state \cite{shannon06}.


The Hamiltonian in the $\delta S^z=1$ sector is simply a bosonic hopping Hamiltonian and can
be solved exactly.
For $\sin(\theta)>0$ the lowest energy state with  $\delta S^z=1$ has an energy gap
$\Delta(\delta S^z=1)=J (-3 \cos(\theta)-\sin(\theta))$.


In Fig. \ref{fig:boundstate}, this energy is compared with the lowest energy
state of the $\delta S^z=2$  sector, calculated using ED
on a 128-site cubic cluster with
periodic boundary conditions.
Starting from the AIAO phase and approaching the boundary with the proposed
nematic QSL we see that the energy of the $\delta S^z=2$ sector
comes below the energy of the $\delta S^z=1$ sector.
This indicates the formation of a two--magnon bound state with lower energy
than the lowest single--magnon state.
The two--magnon bound state crosses the AIAO state  at $\theta \approx 1.92$,
indicating an instability to nematic order.
This is in good agreement with the phase boundary $\theta \approx 1.93$ 
found using cVAR [Fig.~\ref{fig:phase.diagram}(a)].


\begin{figure}
\centering
\includegraphics[width=0.47\columnwidth]{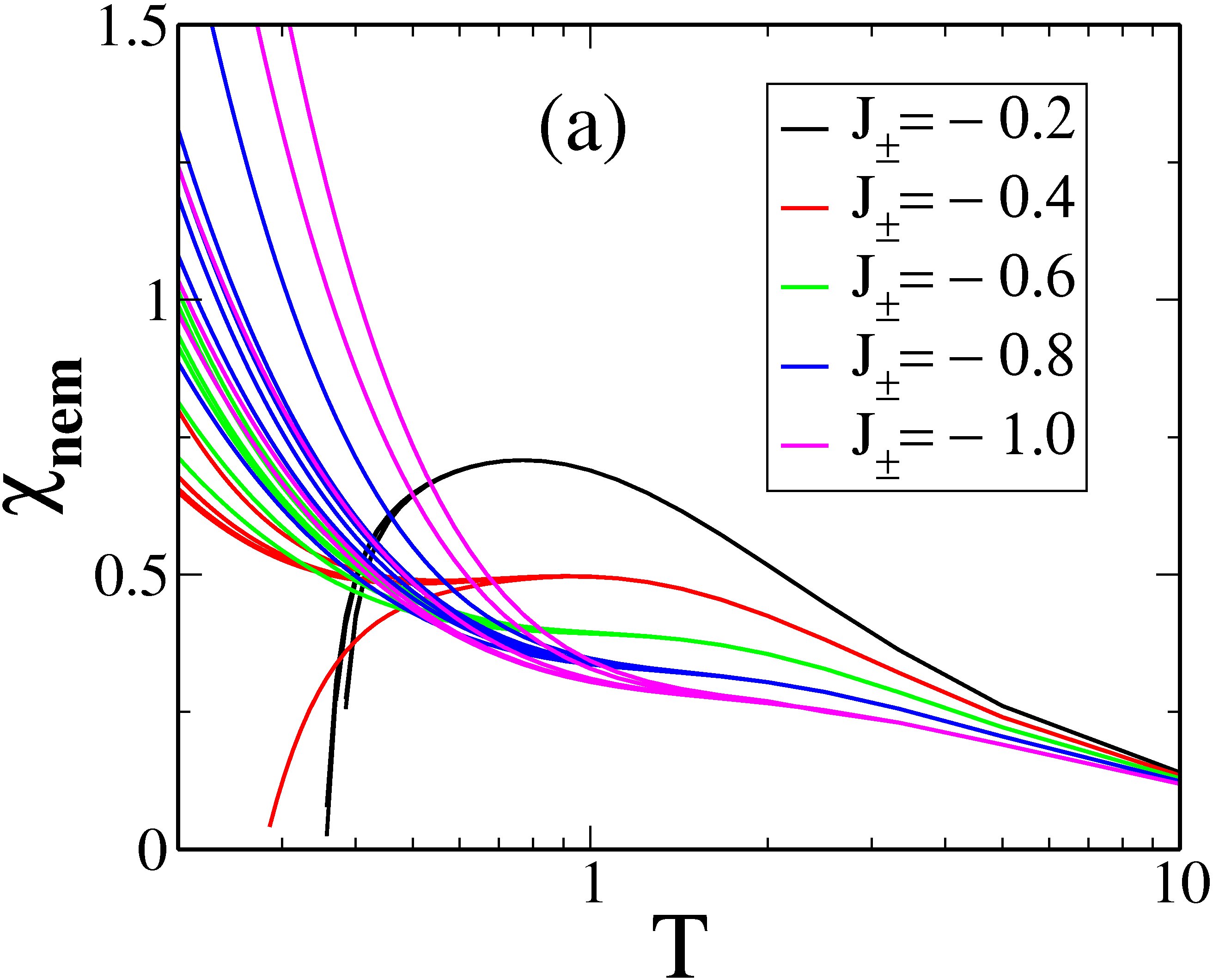}
\includegraphics[width=0.45\columnwidth]{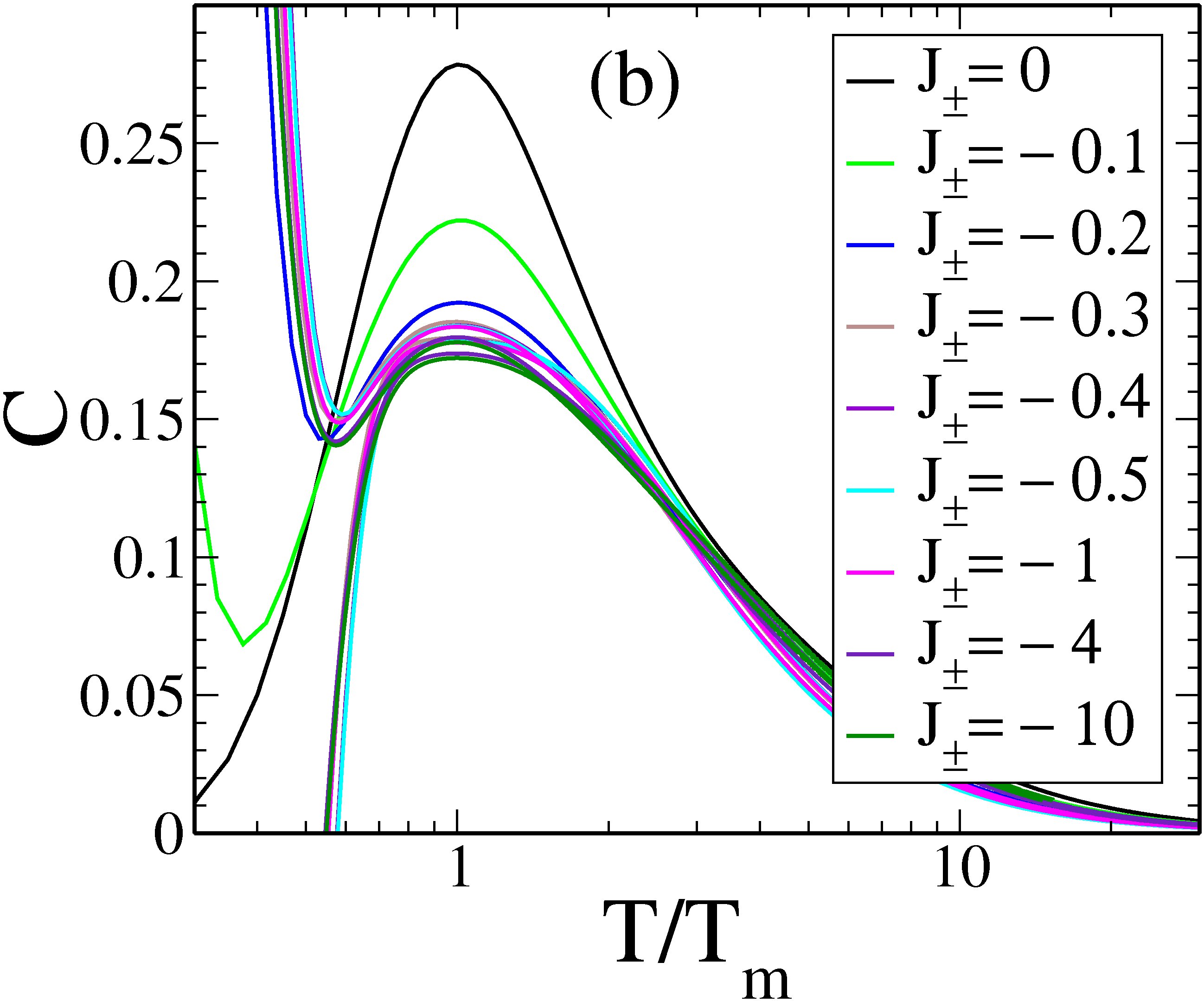}
\caption{
\label{fig:thermodynamics}
Finite--temperature properties of frustrated quantum 
spin ice [Eq.~\ref{eq:Hxxz}], calculated for $J_{zz} = 1$.  
(a) Susceptibilty $\chi_{\sf nem}(T)$ associated with spin--nematic 
order [Eq.~(\ref{eq:Qperp})], 
calculated using high--temperature series expansion (HTE).
Different curves for a given value of $J_\pm$ 
correspond to different Pad{\'e} approximants.
For $J_\pm < -0.5$, 
$\chi_{\sf nem}(T)$ shows an upturn at low temperatures, 
consistent with an approach to spin--nematic order.
(b) Heat capacity $C$, as a function of reduced 
temperature $T/T_m$, calculated within a 
numerical linked--cluster expansion (NLCE). 
Here $T_m$ reflects the temperature at which the heat 
capacity takes on its maximum value for a given value of $J_\pm$;
different curves for the same value of $J_\pm$ represent different
orders of NLCE; agreement between these indicates convergence.
Plots of $C(T/T_m)$ for different values of $J_\pm$ collapse onto 
one another for $J_\pm \lesssim  -0.3$.
This is consistent with an extended regime 
where finite--temperature properties are controlled by the 
zero--temperature $SU(2)$ point, $J_\pm =  -0.5$ (equivalently, $\theta = \frac{\pi}{4}$), 
reminiscent of quantum criticality.
}
\end{figure}


{\it Finite temperature}-- Thus far, we have presented evidence
for a $U(1)$ QSL phase with nematic order in the regime of strong,
frustrated transverse exchange in the $T=0$ phase diagram
Fig.~\ref{fig:phase.diagram}(a).
We expect that this nematic order will only manifest itself
at very low temperatures.
In MC simulations of the corresponding classical model,
nematic order arises at temperatures $T\sim10^{-2} J$ [Fig. \ref{fig:phase.diagram}(b)].
This is similar to the energy scale of collinear ground state selection in
CMFT, suggesting a comparable nematic transition temperature in the quantum
model.
This raises the question of what the physics of a spin ice with
strong, frustrated transverse exchange should be like at intermediate temperatures
$T\sim J$.


To address the physics at these intermediate 
temperatures, we turn to series expansion methods.
Specifically we use HTE
\cite{oitmaa-book, oitmaa13, jaubert15} 
and NLCE \cite{jaubert15, applegate12, tang13} to calculate 
the susceptibility $\chi_{\sf nem}(T)$ of the nematic
order parameter [Eq.~(\ref{eq:Qperp})] and the heat capacity $C(T)$.
We focus on the region near the $SU(2)$  point $\theta=\frac{\pi}{4}$,
where our theory predicts a zero-temperature phase transition between
QSI$_\pi$ and spin nematic phases.
This point has been studied recently
using diagrammatic Monte Carlo \cite{huang16}, finding  spin
correlations similar to spin ice down to $T=J/6$, consistent with our cVAR results.


The HTE of the 
nematic susceptibility $\chi_{\sf nem} (T)$ is plotted in Fig. \ref{fig:thermodynamics}(a),
for various values of $\theta$.
HTE converges down to temperatures $T\sim J$, which is not low enough
to see any definitive signature of the onset of nematic order.
However, there is a hint of a zero--temperature phase transition
at $\theta=\frac{\pi}{4}$ in the behaviour of Pad{\'e} approximants of
$\chi_{\sf nem}(T)$ around the $SU(2)$ point. 
For $\theta\lesssim\frac{\pi}{4}$ the  Pad{\'e} approximants  indicate a
suppression of the nematic susceptibility below $T\sim J$, whereas for $\theta>\frac{\pi}{4}$ they
show an upturn at low temperatures.


A further hint of interesting physics at the $SU(2)$  point is revealed in
NLCE calculations of the heat capacity [Fig.~\ref{fig:thermodynamics}(b)].
The calculations show a broad maximum at temperatures $T_m(\theta)$ just above
the temperature at which NLCE fails to converge.
For a wide range of parameters around the $SU(2)$  point, the heat 
capacity curves for different values of $\theta$ can be collapsed onto one another 
by rescaling the temperature axis by $T_m(\theta)$.


This suggests a region of parameter space where the finite--temperature 
physics is controlled by a single point on the zero--temperature phase diagram.
This is reminiscent of quantum criticality, and is consistent with the
scenario of a zero--temperature phase transition between nematic
and QSI$_\pi$ phases at $\theta=\frac{\pi}{4}$.
Further details of HTE and NLCE can be found in the Supplementary Materials.

{\it Conclusions}-- In this Letter we have explored the ground--state 
properties of a minimal model of a ``quantum spin ice'', the spin--1/2 XXZ 
model on the pyrochlore lattice ${\mathcal H}_{\sf xxz}$~[Eq.~(\ref{eq:Hxxz})], 
in the case of frustrated transverse exchange \mbox{$J_\pm < 0$}. 
First, we have determined the ground--state phase diagram 
of this model within a variational approach, cVAR, which builds 
upon the degenerate wave functions found in cluster mean field theory 
(CMFT) [Fig.~\ref{fig:phase.diagram}(a)].
We find that a $U(1)$ QSL derived from spin ice, 
QSI$_\pi$, transforms into another $U(1)$ QSL with easy--plane 
character and hidden spin--nematic order, QSN$_\perp$, at the 
high--symmetry point, \mbox{$J_\pm = -J_{zz}/2$}.   
Further evidence for this quantum phase transition is taken 
from an exact, variational argument; an analysis of 
the two--magnon instability of the neighbouring all--in, all--out ordered 
phase (AIAO) [Fig.~\ref{fig:boundstate}]; and the scaling of thermodynamic 
properties at finite temperature [Fig.~\ref{fig:thermodynamics}].
The results for the quantum ground state are also consistent 
with classical Monte Carlo simulations carried out at finite temperature 
[Fig.~\ref{fig:phase.diagram}(b)], previously discussed 
in \cite{taillefumier17}.

These results offer a rare glimpse 
into the ground--state properties of a highly--frustrated, 
three--dimensional quantum magnet, which is also frustrated 
in the sense of the QMC sign problem.
The variational approach introduced, cVAR, is quite general, and 
could be applied to other frustrated quantum models.
And the fact that the XXZ model on the pyrochlore lattice  can support 
three distinct forms of $U(1)$ QSL, with two of them linked by a point
with $SU(2)$  symmetry, presents a range of new possibilities.
In particular, the nematic QSL, QSN$_\perp$, owns both the 
gauge degrees of freedom and topological excitations of a 
$U(1)$ QSL \cite{hermele04,benton12,lee12,chen16,huang-arXiv}, 
and the Goldstone modes associated with broken spin--rotation 
symmetry [cf. \onlinecite{smerald13,taillefumier17}].  
Exactly how these excitations combine is an interesting, and 
challenging, open problem.

The results also open some interesting new perspectives for experiment.  
Among the most promising candidates for the realization of a 
quantum spin ice are pyrochlore magnets based 
on Pr$^{3+}$ ions \cite{Kimura2013, wen17, Anand16a,sibille-arXiv}.
Our work is particularly relevant to this case, since
microscopic estimates of the transverse exchange interactions in Pr pyrochlores
have found them to be of frustrated sign \cite{Onoda2011a}.
In the light of this, Pr pyrochlores may be proximate 
to the nematic QSL, QSN$_\perp$, which competes 
with QSI$_\pi$ for sufficiently strong transverse exchange.
We anticipate that this phase would present through its gapped, 
and gapless excitations; 
through the fractionalisation of translation symmetry \cite{lee12,chen16}; 
and through the presence of pinch points in quasi--elastic neutron 
scattering \cite{taillefumier17}, which would be expected to ``wash out'' 
at low temperatures \cite{benton12}.
We should note however, that the experimental situation is 
complicated by the role of disorder, which opens up 
new routes to both QSL and non--QSL ground states 
\cite{yaouanc11a,ross12,taniguchi13,savary17-PRL118,petit16-PRB94,martin17,benton-arXiv}.

Other pyrochlores, such as Ce$_2$Sn$_2$O$_7$ \cite{sibille15}, 
have also been identified as QSL candidates, although at present 
the sign of the transverse exchange is unknown.
Given the developing experimental situation, with new 
pyrochlores continuing to be synthesized and characterized \cite{wiebe15},
we are hopeful that a physical realization of a nematic QSL may 
not be too far in the future.


{\it Acknowledgments:} 
The authors are grateful to Judit Romh\'anyi for a careful reading of the manuscript.
This work was supported by the Theory of Quantum Matter Unit of the 
Okinawa Institute of Science and Technology Graduate University (OIST), and by 
the IdEx Bordeaux BIS--Helpdesk (L.J.).
The work of RRPS is supported in part by US National Science 
Foundation grant number DMR--1306048.
O.B. and L.J. acknowledge the hospitality of OIST, where part of this work was 
completed.

\bibliography{prl_draft}

\onecolumngrid

\section*{Supplemental Material}

\section{Classical Monte Carlo Simulations}

The classical phase diagram of Fig.~1(b) has been obtained via Monte Carlo simulations 
of O(3) spins $\vec {\sf S}_{i}$ of length $|\vec {\sf S}_{i}|=1/2$. 
The simulations are based on the heatbath algorithm, using overrelaxation and 
parrallel tempering to facilitate thermalisation. 
A typical run is made of 201 jobs in parallel, each job corresponding 
to a given temperature. 
The values of the temperatures are split on a logarithmic scale from 
$T/J=10^{-3}$ to $T/J=10$. 
Thermalisation takes place in two steps; first a slow annealing from high 
temperature to the temperature of measurement $T$ during $10^{5}$ 
Monte Carlo steps (MCs), followed by thermalisation at temperature 
$T$ during another $10^{5}$ MCs. 
Then, measurements are made every 10 
MCs during $10^{6}$ MCs. 
The system size is $N=8192$ spins ($8\times 8\times 8$ cubic unit cells).

The phase diagram has been obtained using the same recipe 
as in Ref. [\onlinecite{taillefumier17}], which we shall 
briefly summarise here. 
We refer the interested reader to Ref. [\onlinecite{taillefumier17}] for more details.

The transition temperatures are determined by the singularity in the heat capacity. 
The crossover into the spin-ice regime is also conveniently demarcated by a 
broad peak in the heat capacity. 
However, the entropy loss into the other spin liquids is much less vivid and 
we cannot rely on heat-capacity signatures to determine their boundaries.

On the other hand, the three spin liquids (pHAF, SL$_{\perp}$ and SL') 
contain ferromagnetic fluctuations. Let $m$ be the magnetisation of the system. 
The reduced susceptibility, $\chi T\equiv N(\langle m^{2}\rangle -\langle |m|\rangle^{2})$ 
measures the build up of ferromagnetic correlations. 
$\chi T$ thus takes a different value as the system is cooled down into one of 
the spin liquids, with a characteristic point of inflexion between the paramagnetic 
and spin-liquid values. 
We use this point of inflexion, on a logarithmic temperature scale, 
as the qualitative position of the crossover between the paramagnetic 
and spin-liquid regimes.

The pHAF (resp. SL') regime is born from the enhancement of symmetry of 
the Hamiltonian when the easy-plane spin liquid SL$_{\perp}$ meets spin ice 
(resp. AIAO order). 
Hence, the pHAF (resp. SL') vanishes when spin ice (resp. AIAO) 
correlations vanish, giving rise to SL$_{\perp}$ upon cooling. 
In other words, the reduced susceptibility of the corresponding 
order parameters, $\chi_{\rm ice} T$ and $\chi_{\rm AIAO} T$, 
decreases towards zero upon cooling. 
We fix the crossover temperature between pHAF (resp. SL') and 
SL$_{\perp}$ when $\chi_{\rm ice} T$ (resp. $\chi_{\rm AIAO} T$) 
becomes smaller than its high-temperature limit. 
Please note that the AIAO order parameter is
\begin{eqnarray}
m_{\rm AIAO}=\bigg\langle\frac{1}{N}\sum_{i}{\sf S}^z_i \bigg\rangle.
\label{eq:mAIAO}
\end{eqnarray}
This is the order parameter transforming according to the
${\sf A}_2$ representation of the point group, as identified in Refs.
\cite{taillefumier17, yan17}.

\section{Cluster--variational calculation (${\text c}$VAR)}

Here we introduce the cluster--variational (cVAR)
method used to find the $T=0$ quantum phase diagram presented 
in Fig.~1(a) of the main text.
This is an extension of the standard cluster mean field 
theory (CMFT), to a family of variational wave functions which 
can describe states with long--range entanglement.
As such, cVAR provides a variational approach to the 
quantum spin liquids found in frustrated quantum spin ice.
In what follows, we calculate the relevant variational 
parameters peturbatively, reproducing known results for 
the zero-- and $\pi$--flux phases of quantum spin ice
(QSI$_0$ and QSI$_\pi$), 
and allowing us to identify the phase QSN$_\perp$ as
a nematic quantum spin liquid, with $U(1)$ gauge structure.


We begin by reviewing CMFT, which will provide the basis 
of states used to build the cVAR wave function.
The CMFT ground state wavefunction $|\psi^{\sf CMFT} \{ {\bf h} \}\rangle$ is a product over
`A' tetrahedra of single tetrahedron wavefunctions 
$|\phi_t ( \{ \mathbf{h} \})\rangle$
\begin{eqnarray}
|\psi^{\sf CMFT} \{ {\bf h} \}\rangle = \prod_{t \in A} |\phi_t ( \{ \mathbf{h} \})\rangle.
\label{eq:psi_cmft}
\end{eqnarray}
The single tetrahedron wavefunctions $|\phi_t ( \{ \mathbf{h} \})\rangle$
are the ground states of an auxiliary Hamiltonian $\mathcal{H}'(t)$,
defined on each `A' tetrahedron $t$
\begin{eqnarray}
&&\mathcal{H}'(t)=\sum_{\langle ij \rangle \in t} J \bigg[ \cos(\theta) S^z_i S^z_j +
\sin(\theta) (S^x_i S^x_j +S^y_i S^y_j ) \bigg]
-\sum_{i} {\bf h}_i \cdot {\sf S}_i 
\label{eq:hprime}
\\
&&\mathcal{H}'(t)  |\phi_t ( \{ \mathbf{h} \})\rangle= \epsilon_t  |\phi_t ( \{ \mathbf{h} \})\rangle
\label{eq:phi_t}
\end{eqnarray}
and the external fields ${\bf h}_i$ appearing in Eq. (\ref{eq:hprime}) are variational
parameters, chosen to optimize the variational energy
\begin{eqnarray}
E^{\sf CMFT}_{\sf var}= \langle \psi^{\sf CMFT} \{ {\bf h} \}| \mathcal{H}_{\sf XXZ} |\psi^{\sf CMFT} \{ {\bf h} \}\rangle.
\label{eq:evar_cmft}
\end{eqnarray}

The fields ${\bf h}_i$ are classical vectors defined on each site $i$ of the pyrochlore
lattice and can be used to unambiguously index a CMFT wavefunction $|\psi^{\sf CMFT} \{ {\bf h} \}\rangle$,
up to a global complex phase, via Eqs. (\ref{eq:psi_cmft})-(\ref{eq:phi_t}).
Here we are using the notation $\{ {\bf h} \}$ to denote a configuration of fields ${\bf h}_i$
across the whole lattice.

In the AIAO and AF$_\perp$ phases shown in Fig. 1(a) of the main text, the optimal configuration
of ${\bf h}_i$ is unique up to global symmetry operations.
In the AIAO phase, ${\bf h}_i$ is uniform and points along the ${\bf z}$ direction of spin space
\begin{eqnarray}
{\bf h}_i = h \hat{\bf z}, \quad \forall i
\end{eqnarray}
In the AF$_\perp$,  ${\bf h}_i$ is uniform and lies in the ${xy}$ plane of spin space, e.g.
\begin{eqnarray}
{\bf h}_i = h \hat{\bf x}, \quad \forall i
\end{eqnarray}
In these, non-degenerate, cases we do not go beyond standard CMFT.


cVAR is useful in cases where the optimal configuration of ${\bf h}_i$, obtained
in standard CMFT, is highly degenerate.
This occurs, for example, in the regions of parameter space spanned by the 
phases QSI$_0$ and QSI$_\pi$ [cf. Fig. 1(a) of the main text].
Here the solutions for ${\bf h}_i$ found in CFMT comprise an extensive 
set of ``spin ice'' configurations, in which the classical field ${\bf h}$ obeys 
the Bernal--Fowler ice rules.
The same is also true of the CMFT solutions for the nematic quantum 
spin liquid (QSN$_\perp$).
However in this case the classical field ${\bf h}$ obey the more general 
``lego--brick rules'', set out in [\onlinecite{yan17}].



cVAR consists in writing down a new variational wavefunction which is
a superposition of the highly degenerate CMFT wavefunctions.
Each CMFT wavefunction can be unambiguously labelled by
a configuration of fields $\{ {\bf h }\}$.
\begin{eqnarray}
| \varphi^{\sf cVAR} \rangle = \sum_{\{ {\bf h} \}} a_{\{ {\bf h}\}} |\psi^{\sf CMFT} ( \{ {\bf h }  \} ) \rangle
\label{eq:phi_ECMFT}
\end{eqnarray}
and 
the sum runs over all field configurations  $ \{ {\bf h } \}$ which optimize Eq.(\ref{eq:evar_cmft}).
The complex coefficients $a_{\{ {\bf h}\}}$ are new variational parameters, chosen such that
\begin{eqnarray}
\sum_{\{ {\bf h} \}} | a_{\{ {\bf h}\}} |^2=1
\label{eq:anorm}
\end{eqnarray}


These are chosen to optimize the new variational energy, evaluated with the respect to 
the original Hamiltonian $\mathcal{H}_{\sf XXZ}$
\begin{eqnarray}
E_{\sf cVAR}= \frac{\langle\varphi^{\sf cVAR} | \mathcal{H}_{\sf XXZ} |\varphi^{\sf cVAR} \rangle}
{\langle \varphi^{\sf cVAR}  |\varphi^{\sf cVAR} \rangle}.
\label{eq:evar_ecmft}
\end{eqnarray}
The factor of $\langle \varphi^{\sf cVAR}  |\varphi^{\sf cVAR} \rangle$ in the
denominator of Eq. (\ref{eq:evar_ecmft}) is necessary despite Eq. (\ref{eq:anorm}), because the
CMFT wavefunctions $|\psi^{\sf CMFT} ( \{ {\bf h }  \} ) \rangle$ are not necessarily orthogonal.


The cVAR wavefunction [Eq. (\ref{eq:phi_ECMFT})] is able to describe highly--entangled 
phases, such as quantum spin liquids, which could not have been described at the
standard CMFT level.
In general, optimizing $E_{\sf cVAR}$ [Eq. (\ref{eq:evar_ecmft})], 
would require a sophisticated variational Monte Carlo calculation. 
However in each of the cases considered here, we are able to use a perturbative 
expansion of the cVAR energy to map the problem 
onto a previously--solved model of a U(1) QSL.   
We will first illustrate the cVAR procedure for the QSI$_0$ and QSI$_{\pi}$
regions of the phase diagram, showing how it produces agreement with previously
established results from other methods.
We will then demonstrate its application to the QSN$_{\perp}$ phase.
\\

\subsection{${\text c}$VAR for QSI$_0$ and QSI$_{\pi}$}

At the level of CMFT, we cannot distinguish between the QSI$_0$ and QSI$_{\pi}$
regions of the phase diagram.
Throughout the region of parameter space spanned by the QSI$_0$ and QSI$_{\pi}$
the optimal configurations of ${\bf h}_i$ found in CMFT are of the form
\begin{eqnarray}
{\bf h}_i= \sigma_i h \hat{{\bf z}}_i , \quad \sigma_i=\pm1
\label{eq:SI_cmft}
\end{eqnarray}
where the sign factors $\sigma_i$ obey an ``ice rule'' constraint,
summing to zero on every tetrahedron (both `A' and `B' tetrahedra) of 
the lattice
\begin{eqnarray}
\sum_{i \in t} \sigma_i=0 , \quad \forall \ \ 
\text{tetrahedra} \ t.
\label{eq:icerule_cmft}
\end{eqnarray}
The field strength $h$ is uniform and determined by the
optimization of the CMFT energy Eq. (\ref{eq:evar_cmft}).
The arrangement of sign variables $\sigma_i$ is thus the only thing
distinguishing degenerate mean field solutions.

We therefore label mean field solutions by the sign configuration $\{ \sigma \}$
and define
\begin{eqnarray}
&&|\psi^{\sf CMFT} ( \{ \sigma \}) \rangle \equiv
|\psi^{\sf CMFT} ( \{ \sigma h \hat{\bf z} \}) \rangle \\
&&a_{\{ \sigma \}}  \equiv
a_{\{ \sigma h {\bf z} \}} 
\end{eqnarray}
cf. Eqs. (\ref{eq:psi_cmft}), (\ref{eq:phi_t}) and (\ref{eq:phi_ECMFT}).

We now wish to consider a superposition of CMFT solutions of the
form of Eq. (\ref{eq:phi_ECMFT}), and the associated variational energy
Eq. (\ref{eq:evar_ecmft}).
In order to evaluate Eq. (\ref{eq:evar_ecmft}) we need to calculate both
the overlap 
\begin{eqnarray}
O_{\{ \sigma' \} \{\sigma\}} = 
\langle \psi^{\sf CMFT} ( \{ \sigma'  \} ) |\psi^{\sf CMFT} ( \{\sigma  \} ) \rangle
\label{eq:Odef}
\end{eqnarray}
and the Hamiltonian matrix element 
\begin{eqnarray}
X_{\{ \sigma' \} \{ \sigma \}}=
\langle \psi^{\sf CMFT} ( \{ \sigma'  \} ) | \mathcal{H}_{\sf XXZ} |\psi^{\sf CMFT} ( \{ \sigma  \} ) \rangle
\label{eq:Xdef}
\end{eqnarray}
between a general pair of CMFT wavefunctions, labelled by field configurations
$ \{ {\bf h} \}$ and $ \{ {\bf h}' \}$, respecting Eqs. (\ref{eq:SI_cmft})-(\ref{eq:icerule_cmft}).
In terms of these properties, the variational energy [Eq. (\ref{eq:evar_ecmft})] becomes
\begin{eqnarray}
E_{\sf var}^{\sf cVAR}=\frac{E_{\sf var}^{\sf CMFT}+\sum_{\{ \sigma \} \neq \{\sigma'\}} 
X_{\{ \sigma' \} \{ \sigma \}}
a_{ \{ \sigma'\} }^{\ast} a^{\phantom \ast}_{ \{\sigma\}}}
{1 + \sum_{\{ \sigma \} \neq \{ \sigma'\}} 
O_{\{ \sigma' \} \{\sigma\}}
a_{ \{ \sigma'\} }^{\ast} a^{\phantom \ast}_{ \{\sigma\}}}
\label{eq:evar_OX}
\end{eqnarray}

\begin{figure}
\centering
\includegraphics[width=0.4\textwidth]{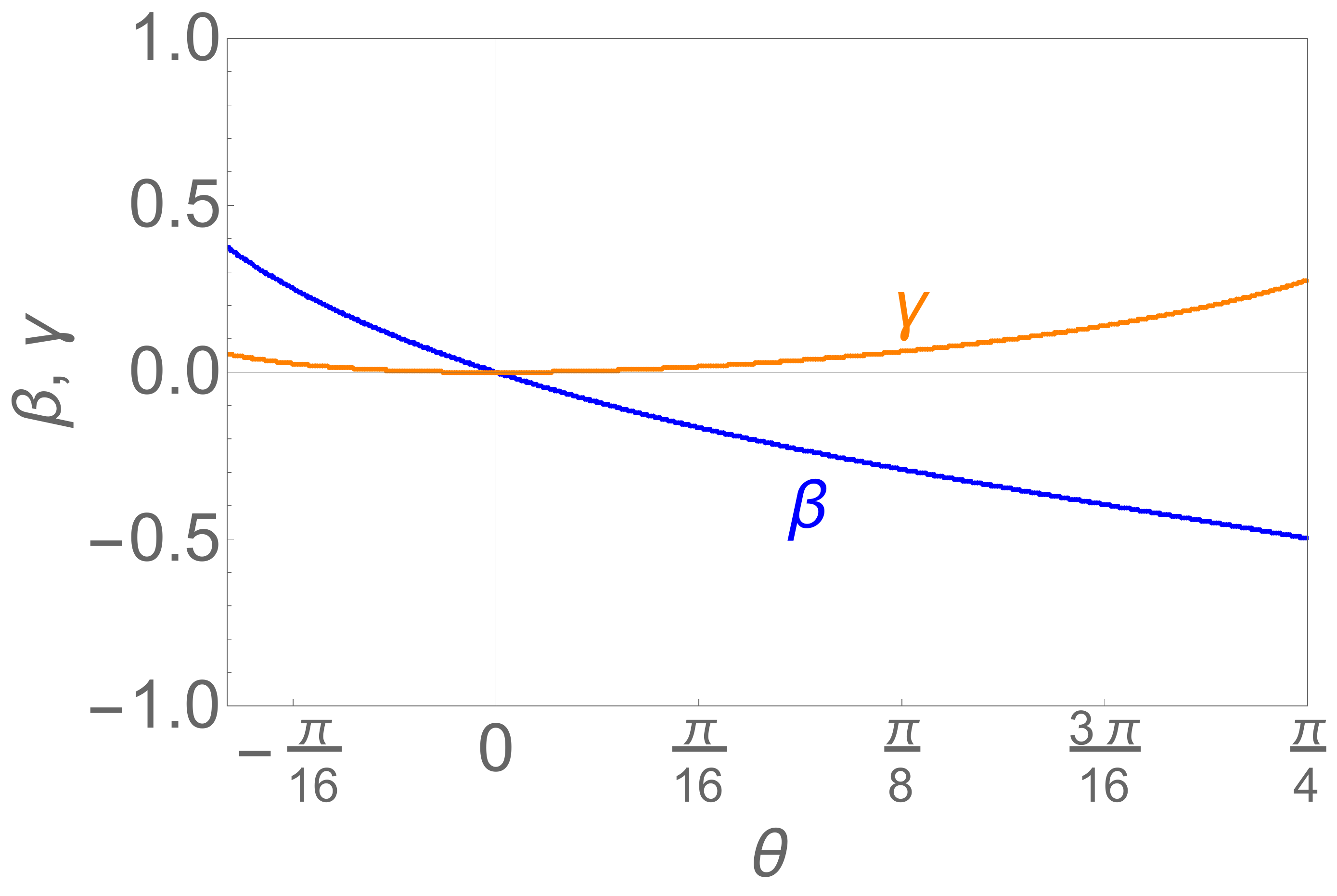}
\caption{Wavefunction parameters $\beta$ and $\gamma$ [see Eq. (\ref{eq:wf_cmft})]
determined in CMFT as a function of the exchange parameter $\theta$, in the region
of parameter space corresponding to the QSI$_0$ and QSI$_{\pi}$ phases.
These determine the overlap and Hamiltonian matrix elements between different
CMFT ground states and function as the small parameters for the expansion of
the cVAR variational energy.
}
\label{fig:qsi_wf}
\end{figure}

In order to calculate $O_{\{ \sigma' \} \{ \sigma \}}$ and $X_{\{ \sigma' \} \{ \sigma \}}$
we need to know how the single tetrahedron wavefunctions $|\phi_t (\{ {\bf h} \})\rangle$ 
[Eqs. (\ref{eq:psi_cmft})-(\ref{eq:phi_t})]
depend on the field configuration on tetrahedron $t$ of the `A' sublattice.
There are 6 possible forms for $|\phi_t (\{ {\bf h} \})\rangle$, corresponding
to the 6 possible arrangements of the sign factors $\sigma_i=\pm1$ [Eq. (\ref{eq:SI_cmft})] on 
a single tetrahedron.
Labelling each possible  $|\phi_t (\{ {\bf h} \})\rangle$  according to the 
associated arrangement of sign factors (e.g. $|++--\rangle$), and writing them out
in the basis of eigenstates of $S^z_i$ ($|\uparrow\rangle, |\downarrow\rangle$)
we obtain:
\begin{eqnarray}
&&|++--\rangle=\sqrt{1-\beta^2-\gamma^2} | \uparrow \uparrow \downarrow \downarrow \rangle + 
 \frac{\beta}{2}
\bigg( | \uparrow \downarrow \uparrow \downarrow  \rangle
+
 | \uparrow \downarrow \downarrow \uparrow \rangle
+
 |\downarrow \uparrow \uparrow  \downarrow \rangle
+ 
|  \downarrow \uparrow \downarrow \uparrow\rangle
\bigg)
+\gamma  |\downarrow \downarrow\uparrow \uparrow  \rangle
\nonumber \\
&&|+-+-\rangle=
\sqrt{1-\beta^2-\gamma^2}  
| \uparrow \downarrow \uparrow \downarrow  \rangle
+ 
 \frac{\beta}{2}
\bigg( | \uparrow \uparrow \downarrow \downarrow \rangle 
+
 | \uparrow \downarrow \downarrow \uparrow \rangle
+
 |\downarrow \uparrow \uparrow  \downarrow \rangle
+ 
 |\downarrow \downarrow\uparrow \uparrow  \rangle
\bigg)
+\gamma |  \downarrow \uparrow \downarrow \uparrow\rangle
\nonumber \\
&&|+--+\rangle=
\sqrt{1-\beta^2-\gamma^2}  
 | \uparrow \downarrow \downarrow \uparrow \rangle
+ 
 \frac{\beta}{2}
\bigg( | \uparrow \uparrow \downarrow \downarrow \rangle 
+
| \uparrow \downarrow \uparrow \downarrow  \rangle
+
|  \downarrow \uparrow \downarrow \uparrow\rangle
+ 
 |\downarrow \downarrow\uparrow \uparrow  \rangle
\bigg)
+\gamma
 |\downarrow \uparrow \uparrow  \downarrow \rangle 
\nonumber \\
&&|-++-\rangle=
\sqrt{1-\beta^2-\gamma^2}  
 |\downarrow \uparrow \uparrow  \downarrow \rangle 
+ 
 \frac{\beta}{2}
\bigg( | \uparrow \uparrow \downarrow \downarrow \rangle 
+
| \uparrow \downarrow \uparrow \downarrow  \rangle
+
|  \downarrow \uparrow \downarrow \uparrow\rangle
+ 
 |\downarrow \downarrow\uparrow \uparrow  \rangle
\bigg)
+\gamma
 | \uparrow \downarrow \downarrow \uparrow \rangle
\nonumber \\
&&|-+-+\rangle=
\sqrt{1-\beta^2-\gamma^2}  
|  \downarrow \uparrow \downarrow \uparrow\rangle
+ 
 \frac{\beta}{2}
\bigg( | \uparrow \uparrow \downarrow \downarrow \rangle 
+
 | \uparrow \downarrow \downarrow \uparrow \rangle
+
 |\downarrow \uparrow \uparrow  \downarrow \rangle
+ 
 |\downarrow \downarrow\uparrow \uparrow  \rangle
\bigg)
+\gamma 
| \uparrow \downarrow \uparrow \downarrow  \rangle
\nonumber \\
&&|--++\rangle=\sqrt{1-\beta^2-\gamma^2}  
 |\downarrow \downarrow\uparrow \uparrow  \rangle
+ 
 \frac{\beta}{2}
\bigg( | \uparrow \downarrow \uparrow \downarrow  \rangle
+
 | \uparrow \downarrow \downarrow \uparrow \rangle
+
 |\downarrow \uparrow \uparrow  \downarrow \rangle
+ 
|  \downarrow \uparrow \downarrow \uparrow\rangle
\bigg)
+\gamma  
| \uparrow \uparrow \downarrow \downarrow \rangle
\label{eq:wf_cmft}
\end{eqnarray}
where $\beta$ and $\gamma$ are real
functions of the exchange parameters.
These are determined as a function of $\theta$ from
CMFT and are plotted in Fig. \ref{fig:qsi_wf}.

To evaluate $O_{\{ \sigma' \} \{ \sigma \}}$ and $X_{\{ \sigma' \} \{ \sigma \}}$
we need to calculate the overlaps and Hamiltonian matrix elements between
the single tetrahedron wavefunctions.
These are
\begin{eqnarray}
&&\langle +- +- |++--\rangle= \beta \left( \sqrt{1-\beta^2-\gamma^2}+\gamma + \frac{\beta}{2}\right)  
\label{eq:overlap1}
\\
&&\langle --++ |++--\rangle= \beta^2+2 \gamma \sqrt{1-\beta^2-\gamma^2}  \\
&&\langle +- +- | S^z_0 |++--\rangle=\frac{\beta}{2} \left( \sqrt{1-\beta^2-\gamma^2} - \gamma \right) \\
&&\langle -+-+ | S^z_0 |++--\rangle= \langle --++ | S^z_0 |++--\rangle=0\\
&&\langle -+-+ | S^x_0 |++--\rangle= \langle --++ | S^x_0 |++--\rangle=0\\
 \\
&&\langle +- +- | \mathcal{H}_{\sf XXZ}^{(A)} |++--\rangle=  
-\frac{J}{4} \cos(\theta)  \beta \left( \beta + 2 (\gamma + \sqrt{1-\beta^2-\gamma^2}) \right)+
J \sin(\theta)  \left(\frac{1}{2}+ (\beta+\gamma) (\beta+\sqrt{1-\beta^2-\gamma^2})\right) \nonumber \\
 \\
&&\langle --++ | \mathcal{H}_{\sf XXZ}^{(A)} |++--\rangle=
- J \cos(\theta) \left( \frac{\beta^2}{2}+\gamma\sqrt{1-\beta^2-\gamma^2} \right)
+ J \sin(\theta) \beta \left( \beta + 2 (\gamma+\sqrt{1-\beta^2-\gamma^2}) \right)
\label{eq:H3}
\end{eqnarray}
where $\mathcal{H}_{\sf XXZ}^{(A)}$ is the Hamiltonian on the `$A$' tetrahedra.
All the other relevant overlaps and matrix elements can be generated from Eqs.
(\ref{eq:overlap1})-(\ref{eq:H3})  using symmetries of the problem.

Both $\beta$ and $\gamma$ are significantly smaller than 1 over the whole
regime where the optimal CMFT  state is of the form of Eq. (\ref{eq:SI_cmft})
[see Fig. \ref{fig:qsi_wf}].
Using this fact we can expand Eqs. (\ref{eq:overlap1})-(\ref{eq:H3})  to linear order
in $\beta, \gamma$ and obtain
\begin{eqnarray}
&&\langle +- +- |++--\rangle\approx \beta
\\
&&\langle --++ |++--\rangle\approx 2\gamma  \\
&&\langle +- +- | S^z_0 |++--\rangle\approx\frac{\beta}{2}\\
&&\langle +- +- | \mathcal{H}_{\sf XXZ}^{(A)} |++--\rangle \approx
-\frac{J}{2} \cos(\theta)  \beta +
J \sin(\theta)  \left(\frac{1}{2}+ (\beta+\gamma)\right) \nonumber \\
 \\
&&\langle --++ | \mathcal{H}_{\sf XXZ}^{(A)} |++--\rangle\approx
- J \cos(\theta) \gamma
+ 2 J \sin(\theta) \beta 
\end{eqnarray}

\begin{figure}
\centering
\includegraphics[width=0.6\textwidth]{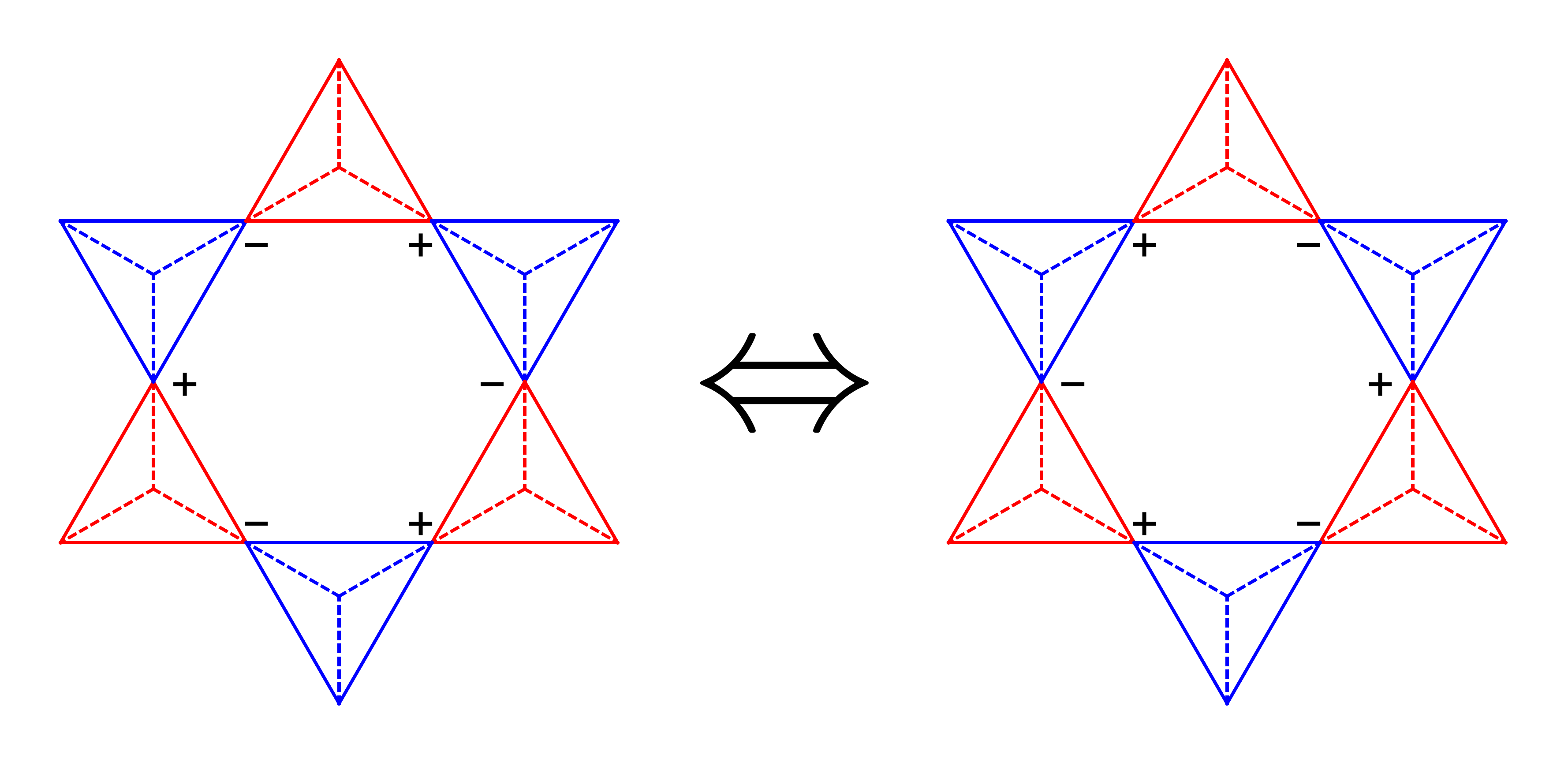}
\caption{CMFT wavefunctions related by flipping the sign variables $\sigma_i=\pm1$ around a single hexagonal
plaquette provided the leading terms in the expansions of 
$X_{\{ \sigma' \}, \{ \sigma \}}$ 
and $O_{\{ \sigma' \}, \{ \sigma \}}$ 
[Eqs. (\ref{eq:Odef}), (\ref{eq:Xdef}), (\ref{eq:Xexpand}), (\ref{eq:Oexpand})].}
\label{fig:hexflip}
\end{figure}

Using these results and Eq. (\ref{eq:psi_cmft}) we can find the leading terms
in $\beta$ and $\gamma$ in the numerator and denominator of Eq. (\ref{eq:evar_OX}).
The leading term in the sums in both numerator and denominator comes from pairs of configurations $\{ \sigma \}$
and $\{ \sigma' \}$ which are related by reversing the sign factors $\sigma_i$ on six sites around a single
hexagonal plaquette [Fig. \ref{fig:hexflip}]. For two such configurations we have
\begin{eqnarray}
&&X_{\{ \sigma' \} \{ \sigma \}}= \frac{3}{2} J \sin(\theta) \beta^2 +\mathcal{O}(\beta^3) 
\label{eq:Xexpand}
\\
&&O_{\{ \sigma' \} \{ \sigma \}}=\beta^3 +\mathcal{O}(\beta^4).
\label{eq:Oexpand}
\end{eqnarray}

Using this to expand Eq.  (\ref{eq:evar_OX}) up to order $\beta^2$, gives a new equation for the
variational energy
\begin{eqnarray}
E_{\sf var}^{\sf cVAR}\approx E_{\sf var}^{\sf CMFT}+\sum_{\{ \sigma \} \neq \{ \sigma'\}} 
M_{\{ \sigma' \} \{ \sigma \}})
a_{\{ \sigma' \}}^{\ast} 
a_{\{ \sigma \}}^{\phantom\ast} 
\label{eq:evar_expanded}
\end{eqnarray}
where
\begin{eqnarray}
M_{\{ \sigma' \} \{ \sigma \}}=\frac{3}{2} J \sin(\theta) \beta^2
\label{eq:matrixelement}
\end{eqnarray}
for two configurations related by flipping a single 
hexagonal plaquette and zero otherwise.

We now face the problem of finding the set of coefficients $a_{\{ {\bf h} \}}$ which will
optimize the expanded variational energy Eq.~(\ref{eq:evar_expanded}), as a function of $\theta$. 
Fortunately, the solution to this problem is already known.

Optimizing the variational energy in  Eq.~(\ref{eq:evar_expanded})
 is equivalent to solving the ring exchange problem studied
by Quantum Monte Carlo in Ref. \cite{shannon12}.
The results tell us that when $M_{\{ \sigma' \} \{ \sigma \}}<0$
the optimum wavefunction is the 0-flux $U(1)$ quantum spin liquid which
we refer to as QSI$_0$.
Since $M_{\{ \sigma' \} \{ \sigma \}}$ has the same sign as 
$\sin(\theta)$, this leads us to assign the region $-0.26<\theta<0$
to the QSI$_0$ phase.

For $\theta>0\implies M_{\{ \sigma' \} \{ \sigma \}}>0 $ we have
the same problem but now with a positive tunnelling matrix element.
Using a gauge transformation described in Refs. \cite{hermele04, lee12}
one can relate this case back to the case with $M_{\{ \sigma' \} \{ \sigma \}}<0$,
and find that the ground state is also a quantum spin liquid but now of the $\pi$-flux
variety ( QSI$_{\pi}$).
We therefore assign the region $0<\theta<\frac{\pi}{4}$ to the  QSI$_{\pi}$ phase.

We therefore identify the phase boundary between
0-flux and $\pi$-flux QSLs
at $\theta=0$, which is in agreement with the known result from 
perturbation theory \cite{hermele04, lee12}.

\subsection{${\text c}$VAR for QSN$_{\perp}$}

Having established the general method, and applied it to distinguish between the 
0-flux and $\pi$-flux QSLs in the region with spin-ice-like CMFT ground states, we now
demonstrate its application for the region of strong frustrated transverse exchange
$\frac{\pi}{4}<\theta\lesssim1.93$.

In this region, the optimal CMFT solutions correspond to field configurations
of the form
\begin{eqnarray}
{\bf h}_i= \sigma_i h \hat{{\bf x}}, \quad \sigma_i=\pm1
\label{eq:nem_cmft}
\end{eqnarray}
and those related to Eq. (\ref{eq:nem_cmft})
by global symmetry transformations.
The choice of a global axis within the $xy$ plane
for ${\bf h}_i$ indicates the spontaneous breaking of
$U(1)$ spin rotation symmetry.
The sign variables $ \sigma_i=\pm1$ can take on any 
one of an extensively large number of configurations obeying
the constraint Eq. (\ref{eq:icerule_cmft}) on every tetrahedron of the
lattice.

We proceed with the cVAR method in precisely the same way as
above: by writing down a new wavefunction which is a superposition
of CMFT solutions [Eq. (\ref{eq:phi_ECMFT})] and seeking to optimize
its variational energy [Eq. (\ref{eq:evar_ecmft})].

We consider a superposition of CMFT solutions 
\begin{eqnarray}
|\psi^{\sf CMFT}(\{ \sigma \}) \rangle =|\psi^{\sf CMFT}(\{ h \sigma \hat{\bf x} \}) \rangle
\end{eqnarray}
with a fixed global axis of collinearity (in this case $\hat{{\bf x}}$).
Pairs of CMFT wavefunctions with different global axes of collinearity have vanishing
overlaps and Hamiltonian matrix elements between them in the thermodynamic limit,
so superposing states with different collinearity axes would not improve the variational energy. 

To write down the single tetrahedron wavefunctions, from which the CMFT wavefunctions are formed via
Eq. (\ref{eq:psi_cmft}), it is convenient to use the basis of eigenstates of ${\sf S}^x_i$ which we write as
$|\rightarrow\rangle, |\leftarrow\rangle$.
As before, there are six possible single tetrahedron wavefunctions, indexed by 6 possible arrangements
of signs $\sigma_i$, constrained by Eq. (\ref{eq:icerule_cmft}).

\begin{eqnarray}
&&|++--\rangle=\sqrt{1-\mu^2-\nu^2-\rho^2} | \rightarrow \rightarrow \leftarrow \leftarrow \rangle + 
 \frac{\mu}{2}
\bigg( | \rightarrow \leftarrow \rightarrow \leftarrow  \rangle
+
 | \rightarrow \leftarrow \leftarrow \rightarrow \rangle
+
 |\leftarrow \rightarrow \rightarrow  \leftarrow \rangle
+ 
|  \leftarrow \rightarrow \leftarrow \rightarrow\rangle
\bigg)
+\nu  |\leftarrow \leftarrow\rightarrow \rightarrow  \rangle
\nonumber \\
&& \qquad \qquad  \qquad
+ \frac{\rho}{\sqrt{2}} (| \leftarrow \leftarrow \leftarrow \leftarrow \rangle +
| \rightarrow \rightarrow \rightarrow \rightarrow \rightarrow  ) \nonumber \\
&&|+-+-\rangle=
\sqrt{1-\mu^2-\nu^2-\rho^2}  
| \rightarrow \leftarrow \rightarrow \leftarrow  \rangle
+ 
 \frac{\mu}{2}
\bigg( | \rightarrow \rightarrow \leftarrow \leftarrow \rangle 
+
 | \rightarrow \leftarrow \leftarrow \rightarrow \rangle
+
 |\leftarrow \rightarrow \rightarrow  \leftarrow \rangle
+ 
 |\leftarrow \leftarrow\rightarrow \rightarrow  \rangle
\bigg)
+\nu |  \leftarrow \rightarrow \leftarrow \rightarrow\rangle
\nonumber \\
&& \qquad \qquad  \qquad
+ \frac{\rho}{\sqrt{2}} (| \leftarrow \leftarrow \leftarrow \leftarrow \rangle +
| \rightarrow \rightarrow \rightarrow \rightarrow \rightarrow  ) \nonumber \\
&&|+--+\rangle=
\sqrt{1-\mu^2-\nu^2-\rho^2}  
 | \rightarrow \leftarrow \leftarrow \rightarrow \rangle
+ 
 \frac{\mu}{2}
\bigg( | \rightarrow \rightarrow \leftarrow \leftarrow \rangle 
+
| \rightarrow \leftarrow \rightarrow \leftarrow  \rangle
+
|  \leftarrow \rightarrow \leftarrow \rightarrow\rangle
+ 
 |\leftarrow \leftarrow\rightarrow \rightarrow  \rangle
\bigg)
+\nu
 |\leftarrow \rightarrow \rightarrow  \leftarrow \rangle 
\nonumber \\
&& \qquad \qquad  \qquad
+ \frac{\rho}{\sqrt{2}} (| \leftarrow \leftarrow \leftarrow \leftarrow \rangle +
| \rightarrow \rightarrow \rightarrow \rightarrow \rightarrow  )  \nonumber\\
&&|-++-\rangle=
\sqrt{1-\mu^2-\nu^2-\rho^2}  
 |\leftarrow \rightarrow \rightarrow  \leftarrow \rangle 
+ 
 \frac{\mu}{2}
\bigg( | \rightarrow \rightarrow \leftarrow \leftarrow \rangle 
+
| \rightarrow \leftarrow \rightarrow \leftarrow  \rangle
+
|  \leftarrow \rightarrow \leftarrow \rightarrow\rangle
+ 
 |\leftarrow \leftarrow\rightarrow \rightarrow  \rangle
\bigg)
+\nu
 | \rightarrow \leftarrow \leftarrow \rightarrow \rangle
\nonumber \\
&& \qquad \qquad  \qquad
+ \frac{\rho}{\sqrt{2}} (| \leftarrow \leftarrow \leftarrow \leftarrow \rangle +
| \rightarrow \rightarrow \rightarrow \rightarrow \rightarrow  ) \nonumber \\ 
&&|-+-+\rangle=
\sqrt{1-\mu^2-\nu^2-\rho^2}  
|  \leftarrow \rightarrow \leftarrow \rightarrow\rangle
+ 
 \frac{\mu}{2}
\bigg( | \rightarrow \rightarrow \leftarrow \leftarrow \rangle 
+
 | \rightarrow \leftarrow \leftarrow \rightarrow \rangle
+
 |\leftarrow \rightarrow \rightarrow  \leftarrow \rangle
+ 
 |\leftarrow \leftarrow\rightarrow \rightarrow  \rangle
\bigg)
+\nu 
| \rightarrow \leftarrow \rightarrow \leftarrow  \rangle
\nonumber \\
&& \qquad \qquad  \qquad
+ \frac{\rho}{\sqrt{2}} (| \leftarrow \leftarrow \leftarrow \leftarrow \rangle +
| \rightarrow \rightarrow \rightarrow \rightarrow \rightarrow  ) \nonumber \\
&&|--++\rangle=\sqrt{1-\mu^2-\nu^2-\rho^2}  
 |\leftarrow \leftarrow\rightarrow \rightarrow  \rangle
+ 
 \frac{\mu}{2}
\bigg( | \rightarrow \leftarrow \rightarrow \leftarrow  \rangle
+
 | \rightarrow \leftarrow \leftarrow \rightarrow \rangle
+
 |\leftarrow \rightarrow \rightarrow  \leftarrow \rangle
+ 
|  \leftarrow \rightarrow \leftarrow \rightarrow\rangle
\bigg)
+\nu  
| \rightarrow \rightarrow \leftarrow \leftarrow \rangle 
\nonumber \\
&& \qquad \qquad  \qquad
+ \frac{\rho}{\sqrt{2}} (| \leftarrow \leftarrow \leftarrow \leftarrow \rangle +
| \rightarrow \rightarrow \rightarrow \rightarrow \rightarrow  )
\label{eq:wf_nem}
\end{eqnarray}

\begin{figure}
\centering
\includegraphics[width=0.4\textwidth]{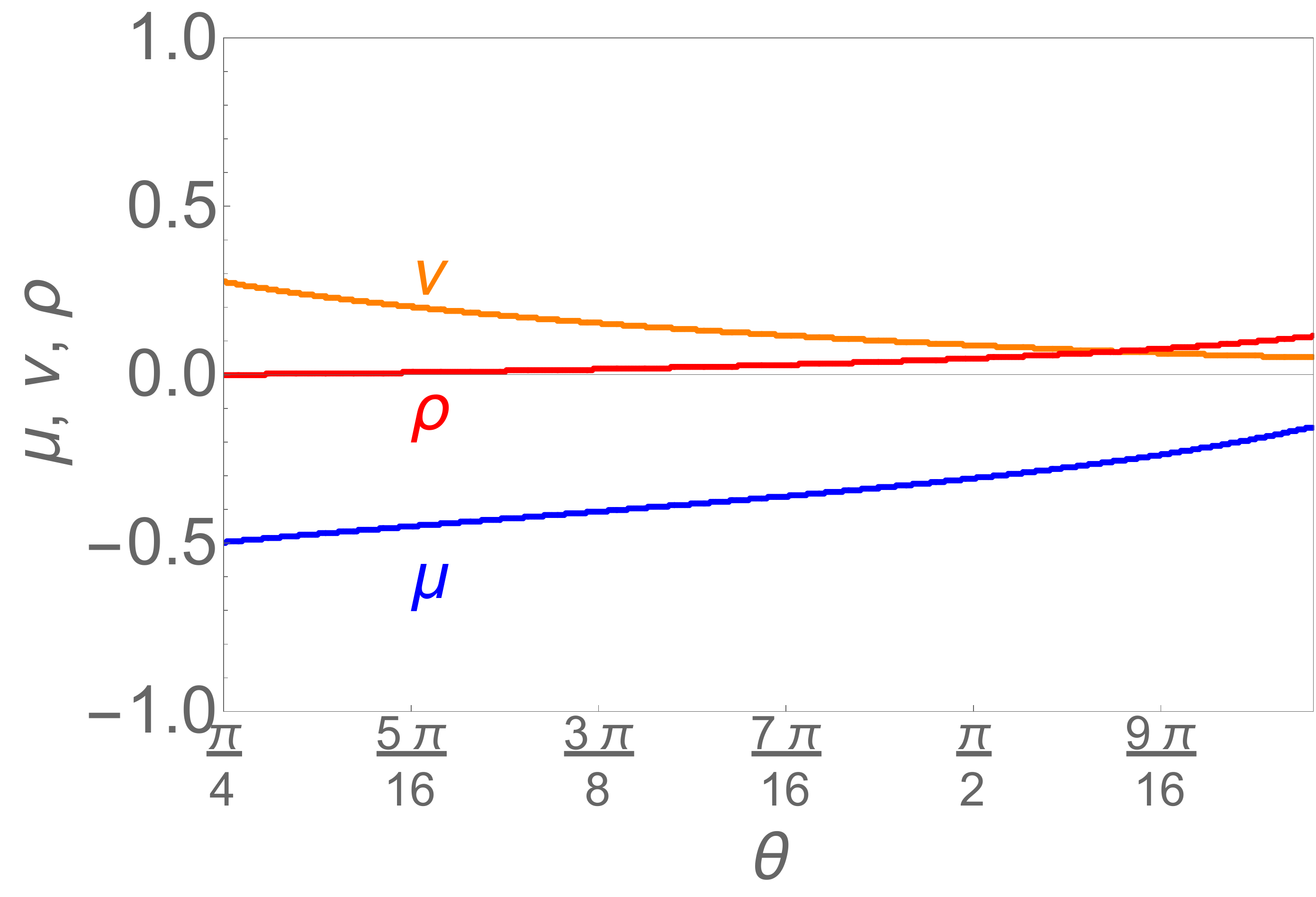}
\caption{Wavefunction parameters $\mu$, $\nu$  and $\rho$ [see Eq. (\ref{eq:wf_nem})]
determined in CMFT as a function of the exchange parameter $\theta$, in the region
of parameter space corresponding to the nematic QSL phase.
These determine the overlap and Hamiltonian matrix elements between different
CMFT ground states and function as the small parameters for the expansion of
the cVAR variational energy.
}
\label{fig:wf_nem}
\end{figure}

The wavefunction parameters $\mu, \nu, \rho$ are plotted 
as a function of $\theta$
in Fig. \ref{fig:wf_nem}.
These remain small throughout the relevant region of parameter
space and we use them as small parameters in an expansion of
the variational energy Eq. \ref{eq:evar_OX}.

Up to linear order in $\mu, \nu, \rho$:
\begin{eqnarray}
&&\langle +- +- |++--\rangle\approx \mu
\\
&&\langle --++ |++--\rangle\approx 2\nu  \\
&&\langle +- +- | S^x_0 |++--\rangle\approx\frac{\mu}{2}\\
&&\langle +- +- | \mathcal{H}_{\sf XXZ}^{(A)} |++--\rangle \approx
\frac{J}{4} (\sin(\theta) +\cos(\theta))  +
\frac{J}{2} (\mu+\nu+\sqrt{2} \rho ) \cos(\theta)+
\frac{J}{2} (\nu-\sqrt{2} \rho ) \sin(\theta)+
\nonumber \\
 \\
&&\langle --++ | \mathcal{H}_{\sf XXZ}^{(A)} |++--\rangle\approx
J\left(\mu \cos(\theta)+ \frac{\rho \cos(\theta)}{\sqrt{2}} + \mu \sin(\theta) -
\nu \sin(\theta) - \frac{\rho \sin(\theta)}{\sqrt{2}} \right).
\end{eqnarray}

From this we can calculate the leading terms in $\mu, \nu, \rho$
in both numerator and denominator of Eq. \ref{eq:evar_OX}.
Once again the leading terms come from pairs of configuratons
$\{ \sigma \}$ and $\{ \sigma' \}$ related by reversing the sign
factors $\sigma_i$ on six sites around a single hexagonal plaquette [Fig. (\ref{fig:hexflip})].
We have
\begin{eqnarray}
&&{X}_{\{ \sigma' \} \{ \sigma \}} \approx \frac{3 J}{4} (\cos(\theta)+\sin(\theta)) \mu^2 + \mathcal{O}(\mu^3) \\
&&{O}_{\{ \sigma' \} \{ \sigma \}} \approx \mu^3 + \mathcal{O}(\mu^4)
\end{eqnarray}

Using this to expand Eq.  (\ref{eq:evar_OX}) up to order $\mu^2$, gives 
\begin{eqnarray}
E_{\sf var}^{\sf cVAR}\approx E_{\sf var}^{\sf CMFT}+\sum_{\{ \sigma \} \neq \{ \sigma'\}} M_{\{ \sigma' \} \{ \sigma \}}
a_{\{ \sigma' \}}^{\ast} 
a_{\{ \sigma \}}^{\phantom\ast} 
\label{eq:evar_expanded2}
\end{eqnarray}
where now
\begin{eqnarray}
M_{\{ \sigma' \} \{ \sigma \}}=\frac{3 J}{4} (\cos(\theta)+\sin(\theta)) \mu^2 + \mathcal{O}(\mu^3) 
\label{eq:matrixelement2}
\end{eqnarray}
for two configurations related by flipping a single hexagonal plaquette and zero otherwise.


Once again, optimizing such a variational energy is equivalent to solving the ring--exchange 
problem studied by Quantum Monte Carlo in Ref. \cite{shannon12}.
It follows that the cVAR solution 
in this case is also a U(1) QSL, but one with finite spin--nematic order,  
since its wavefunction is a superposition of states with the same value of the 
nematic order parameter.
We note that the value of the effective ring-tunnelling is positive throughout the 
relevant region of parameter space, such that the nematic QSL should
have the same $U(1)$ flux pattern, and fractionalization of translational symmetry, as QSI$_{\pi}$.
It differs from that phase, however, by the presence of nematic order.
This can be seen by calculating the nematic order parameter [Eq. (7) of main text]
within the cVAR wavefunction, giving
\begin{eqnarray}
|\mathcal{Q}_{\perp}|=\frac{1}{6}\left[ 1- \left(\frac{\rho}{\sqrt{2}} -\mu\right)+ \mathcal{O}(\mu^2) \right] \; ,
\end{eqnarray}
where the parameters $\rho$ and $\mu$ take on the values shown in Fig.~\ref{fig:wf_nem}.

\section{Series expansion methods}

\subsection{\bf High Temperature Expansions}

High temperature series expansion is a well known method for 
calculating properties of statistical models \cite{oitmaa-book}.
Finite temperature properties (for example $A$) of the models,
 in the thermodynamic limit, are expanded in powers
of the inverse temperature $\beta$. 
\begin{equation}
A(\beta) = a_0 + a_1 \beta + a_2 \beta^2 + a_3 \beta^3 + \ldots 
\end{equation}
The coefficients $a_n$ are calculated up to some maximum order 
$n=N$ and these are used to numerically evaluate
the property $A(\beta)$ at different temperatures. 
For lattice statistical models, with short-range interactions, 
these expansion converge absolutely
at sufficiently high temperatures and provide accurate estimates of 
the properties. 
At lower temperatures, outside the
radius of convergence of the power series, one
can use series extrapolation methods 
(such as Pade and d-log Pade approximants) 
to enhance the range of numerical
convergence.

One efficient way to generate the series coefficients is 
by the Linked Cluster method.
In the Linked Cluster formalism, an extensive property 
$P$ for a large translationally invariant
lattice ${\mathcal L}$ with $N$-sites is expressed as 
a sum over all distinct linked
clusters $c$ as
\begin{equation}
{P(\mathcal L)\over N}= \sum_c L(c) \ W(c).
\end{equation}
Here $L(c)$, called the lattice constant, is the number of 
embeddings of the cluster $c$, per site, in the lattice $\mathcal{L}$.
This is a geometrical property that only
depends on the lattice under consideration and not on 
the statistical model.
The quantity $W(c)$ is called the weight of the cluster. 
It is defined by the recursive relation
\begin{equation}
W(c) = P(c) -\sum_{c'} W(c').
\label{eq:HTEweight}
\end{equation}
Here the sum is over all proper subclusters $c'$ of the cluster $c$. 
The quantity $P(c)$ is the property for the finite cluster.
Thus $W(c)$ is entirely defined by the finite cluster $c$, 
and does not depend on the larger lattice.
If one can calculate the series expansions for small clusters, 
then starting with the smallest cluster, Eq.~(\ref{eq:HTEweight}) can be used to calculate 
the series expansions for the weights of the clusters. 
One can prove that the weight of a cluster with $N$ bonds is of order $\beta^N$. 
Thus, once the weights of all clusters up to size $N$ have been calculated, 
the series expansion for the infinite cluster to order $N$ follows.

We have used the HTE method to calculate the 
logarithm of the partition function $\ln{Z}$ from which
thermodynamic properties such as entropy, specific heat and free energy 
follow. 
In addition, we can apply
a field associated with some order-parameter and by 
calculating the free-energy to second order in that
field we can calculate the static susceptibilities associated with that 
order. 
Here, we have calculated static susceptibilities
associated with various magnetic order parameters as well as for 
the nematic order parameter.

\subsection{\bf Numerical Linked Cluster Expansions}

Numerical Linked Cluster (NLC) method is a systematic way to calculate 
thermodynamic and 
ground state properties of lattice statistical models in 
the thermodynamic limit \cite{rigol06}. 
The method uses the graphical basis of series expansions 
(such as high temperature expansions) to express
model properties as a sum of suitably defined weights 
over all linked clusters. 
Rather than obtain weights $W(c)$ as a power series 
in some variable, NLC uses exact diagonalization to
calculate them numerically. 
The calculations are carried out up to some
maximum cluster size, $n$, also called the 
order of the calculation, providing an estimate for
the property  ($P_n$) in each order. 

The method has the advantage of being non-perturbative, 
of incorporating  exact information at short distances, 
and building the thermodynamic limit into the formalism. 
For some problems, it has proven to be more accurate
than high temperature series expansions.

For the NLC method, it is often useful to 
consider clusters consisting only of 
complete units of an extended size. 
For example, here,
on the pyrochlore lattice consisting of 
corner-sharing tetrahedra, 
it proves useful to only consider clusters that 
consist of complete tetrahedra \cite{applegate12, hayre13, oitmaa13, jaubert15}.
This avoids strong oscillations caused by 
clusters with free ends. 
For the classical spin-ice problem, the first order
NLC in terms of tetrahedra, is equivalent to the 
well-known Pauling approximation and is already 
very accurate down to $T=0$ \cite{singh12}.
This also greatly simplifies the problem of graph counting 
as there are very few clusters of complete tetrahedra 
in each order.

The NLC calculations are limited by one's 
ability to exactly diagonalize finite clusters. 
For a general model of quantum spin-ice a 4th
order NLC calculation, involving sum over weights 
for clusters up to 4 tetrahedra were done \cite{applegate12, hayre13, oitmaa13, jaubert15}. 
The maximum number of sites in
these clusters was 13. Here, for the XXZ 
model of interest, $S_z$ is
a good quantum number. 
This allows one to go to go one further order and 
calculate NLC to 5th order. 
The largest cluster
needed for such a calculation has 16 sites.

Since exact diagonalization of finite clusters 
leads to energy-levels and wave-functions, 
the method is most suitable
for calculating thermodynamic properties 
such as specific heat and entropy and 
various equal-time thermal correlation
functions. 
Frequency dependent properties do not usually 
have a convergent NLC expansion at any temperature.
Static linear response functions can be calculated 
but require numerical differentiation of the free energy
with respect to an applied field. 
This reduces the accuracy of the calculation. 
Here we have used NLC to calculate
thermodynamic properties and the 
thermal expectation values of the 
squares of various order parameters.

When correlations in the system are short-ranged, 
$P_n$ converges rapidly 
with $n$ and provides a highly accurate 
numerical value of the property $P$ in
the thermodynamic limit. 
When correlation lengths begin to exceed the 
sizes of the clusters studied, one
can use sequence extrapolation methods to 
estimate the limit of the sequence $\{P_n\}$. 
We have found it useful to consider Euler transformations 
starting with third order.
This ameliorates some of the 
strong oscillations in $\{P_n\}$ with $n$ and
improves the apparent convergence 
down to slightly lower temperatures.

\end{document}